\documentclass[pre,letterpaper,twocolumn,superscriptaddress,floatfix]{revtex4}
\usepackage{graphicx,psfrag,amsmath,amssymb,amsfonts,bbm,latexsym,color,dcolumn,bm}

\begin{document}

\title{Universal and anomalous behavior in the thermalization \\ 
of strongly interacting harmonically trapped gas mixtures}

\author{Francisco Jauffred}
\affiliation{\mbox{Department of Physics, University of Massachusetts, Boston, MA 02125, USA}}

\author{Roberto Onofrio}
\affiliation{\mbox{Dipartimento di Fisica e Astronomia ``Galileo Galilei'', Universit\`a  di Padova, 
Via Marzolo 8, Padova 35131, Italy}}
\affiliation{\mbox{Department of Physics and Astronomy, Dartmouth College, 6127 Wilder Laboratory, 
Hanover, NH 03755, USA}}

\author{Bala Sundaram}
\affiliation{\mbox{Department of Physics, University of Massachusetts, Boston, MA 02125, USA}}

\begin{abstract}
We report on the dynamics of thermalization by extending a generalization of the Caldeira-Leggett model, 
developed in the context of cold atomic gases confined in a harmonic trap, to higher dimensions. 
Universal characteristics en route to thermalization which appear to be independent of dimensionality 
are highlighted, which additionally suggest a scaling analogous to  turbulent mixing in fluid dynamics. 
We then focus on features dependent on dimensionality, with particular regard to the role of angular 
momentum of the two atomic clouds, having in mind the goal of efficient thermalization between the two species. 
Finally, by considering asymmetry in species numbers, we find that nonlinear interspecies interactions 
provide a mode locking mechanism between the majority and minority species, relevant to recent experiments 
involving Fermi-Bose mixtures in the normal phase. 
\end{abstract}

\maketitle

\section{Introduction}

The study of ultracold atomic gases is a vibrant frontier of atomic physics holding the 
promise of a controllable environment in which to realize model Hamiltonians of interest 
in condensed matter physics~\cite{Bloch,Lewenstein}. The controllability includes, among 
other things, the ability to change the sign and the strength of interaction among the 
atoms, and equally importantly, the effective dimensionality of the system. 
This latter relies on trapping at temperatures low enough with respect to the energy quantum of 
the system in the involved dimension, and allows for the study of phenomena otherwise precluded, 
or more difficult to observe, in the three-dimensional context (for relevant cases see for 
instance~\cite{Mur,Dalmonte,Barmashova}). Consequently, a significant number of studies have 
been devoted to developing methods to optimize atomic cooling~\cite{McKay,Onofrio}. 
In particular, the most widespread technique to achieve deeper degeneracy relies on 
sympathetically cooling fermions via coupling to a bosonic species, which are easier to cool 
via evaporative cooling. Theoretically, trying to describe the microscopic mechanism for the 
thermalization of the two atomic species is more challenging as it appears to fall outside 
the standard descriptions adopted in condensed matter physics or in fluid dynamics. 

Recently, we proposed models which extend the Caldeira-Leggett 
framework~\cite{Magalinskii,Ullersma1,Ullersma2,Ullersma3,Ullersma4,Caldeira1,Caldeira2,Caldeira3} 
to a setting where both species exhibit negligible frequency spreads, mimicking the situation 
typical for trapping and cooling atomic gases~\cite{OnSun}. The model was developed and analysed
in the case of one-dimensional systems, for which a simplified yet rich dynamics occurs. 
However, typical configurations for sympathetic cooling are fully three-dimensional, especially in
the earlier stages of cooling when no degrees of freedom can be considered as frozen. 
Therefore generalization of the model presented in~\cite{OnSun} is deemed necessary to 
describe sympathetic cooling in realistic situations necessary to guide possible optimization of 
experimental protocols. This is particularly important for the full quantum degenerate case involving
Fermi-Bose mixtures as, in one-dimension, identical specific heats are 
expected for both Fermi and Bose gases~\cite{Schonhammer,Mullin}.

In this paper we describe progress made in this direction by analysing the thermalization dynamics 
during the earlier stage of evaporative cooling in which atoms can be still considered as 
nondegenerate gases. The paper is organized as follows. In Section II we introduce the full 
three-dimensional model and metrics which may be used to continuously monitor the thermalization dynamics. 
These are then used to confirm results obtained earlier~\cite{OnSun} in the one-dimensional case 
as well as to explore universal aspects of the 
transition from non-equilibrium to equilibrium dynamics. Analogous results are shown in higher dimensions 
with the initial focus on the isolation of common characteristics which are independent of 
dimensionality, which provide a generic view of the thermalization process. 
Later, in Section III, we turn to additional parameters that play a possible role in
thermalization in higher dimensions and show that their importance is primarily in the early stages.
This naturally leads to the identification of setups for which thermalization may proceed at the 
fastest rate. Finally, in Section IV we discuss unique features of the nonlinear interaction 
between the two clouds when there is asymmetry in the atom numbers. Specifically, in the limit of 
strong interactions, the center of mass oscillation frequency of the minority species is forced 
to lock onto that of the majority species, a situation relevant to observations in recent experiments.
More general considerations and future directions of relevance to experimental trapping 
schemes conclude the paper.

\section{Dimensionally independent aspects of thermalization}

The core idea discussed in Ref.~\cite{OnSun} was the extension of the Caldeira-Leggett 
model~\cite{Magalinskii,Ullersma1,Ullersma2,Ullersma3,Ullersma4,Caldeira1,Caldeira2,Caldeira3}, 
as a microscopic description of thermalization features, to the atomic physics context. 
The comparison to the Caldeira-Leggett model here is to its genuinely classical features, 
{\it i.e.} as a model for an external environment capable of describing the relaxation to thermal 
equilibrium of a target system, for instance a harmonic oscillator or a bistable system. 
This is successfully achieved in the Caldeira-Leggett model by coupling the targeted system in a linear fashion 
to an infinite and delocalized ensemble of harmonic oscillators with a continuous and unbounded spectrum of frequencies. 
These two distinctive features of the  model require accommodation in the trapped atom context where 
there is a privileged point in space, the trap origin (thus breaking translational invariance, 
on which phonon propagation relies in the original Caldeira-Leggett model), and the atoms 
oscillate at a well-defined frequency, instead of a continuum of frequencies. 
A  secondary feature is that the numbers of atoms in a trap are considerably smaller than in a typical 
condensed-matter system, resulting in mesoscopic features that can be related to Caldeira-Leggett models 
with finite size baths~\cite{Taylor,Hanggi,Hasegawa1,Carcaterra,Hasegawa2,Hasegawa3}. 
For the specific case of a monochromatic bath, the Caldeira-Leggett model is recovered if the range of interaction 
is large enough and the interaction strength is appropriately chosen though, as discussed in detail in~\cite{Taylor}, 
thermalization does not occur.

The model developed in Ref~\cite{OnSun} considered a classical one-dimensional Hamiltonian, consisting 
of $N_b$ particles constituting the 'bath' and $N_p$ 'test' particles, of the form
\begin{eqnarray}
H_{\mathrm{tot}} & = & \sum_{m=1}^{N_p} \left( \frac{P_{m}^2}{2M} +  \frac{M \Omega^2 Q_{m}^2}{2} \right) +
\sum_{n=1}^{N_b} \left(  \frac{p_{n}^2}{2m} + \frac{m \omega^2 q_n^2}{2} \right) \nonumber \\  
& &  + \gamma_E \sum_{m=1}^{N_p} \sum_{n=1}^{N_b} \exp \left[-\frac{(q_n-Q_m)^2}{\lambda^2}\right],
\label{hamil}
\end{eqnarray}
which leads to thermalization of the two species under rather general conditions. 
Although in Ref.~\cite{OnSun} a velocity-dependent interaction Hamiltonian was also
considered, here we confine ourselves only to an interaction term acting only in configurational 
space, which results in a considerable simplification of the dynamics and faster computational speed. 
The model is readily extended  to higher dimensions simply by vectorizing the phase-space variables 
though, as we will see later, a number of other considerations accompany this generalization.
 Hamilton's equations corresponding to Eq.~\ref{hamil} are integrated numerically using a variable step 
algorithm which preserves integrals of motion, such as the total energy and the total angular momentum, to machine precision. The
initial conditions are drawn from thermal distributions specified by the starting temperatures of each atomic cloud.
\begin{figure}[t]
\includegraphics[width=0.95\columnwidth, clip=true]{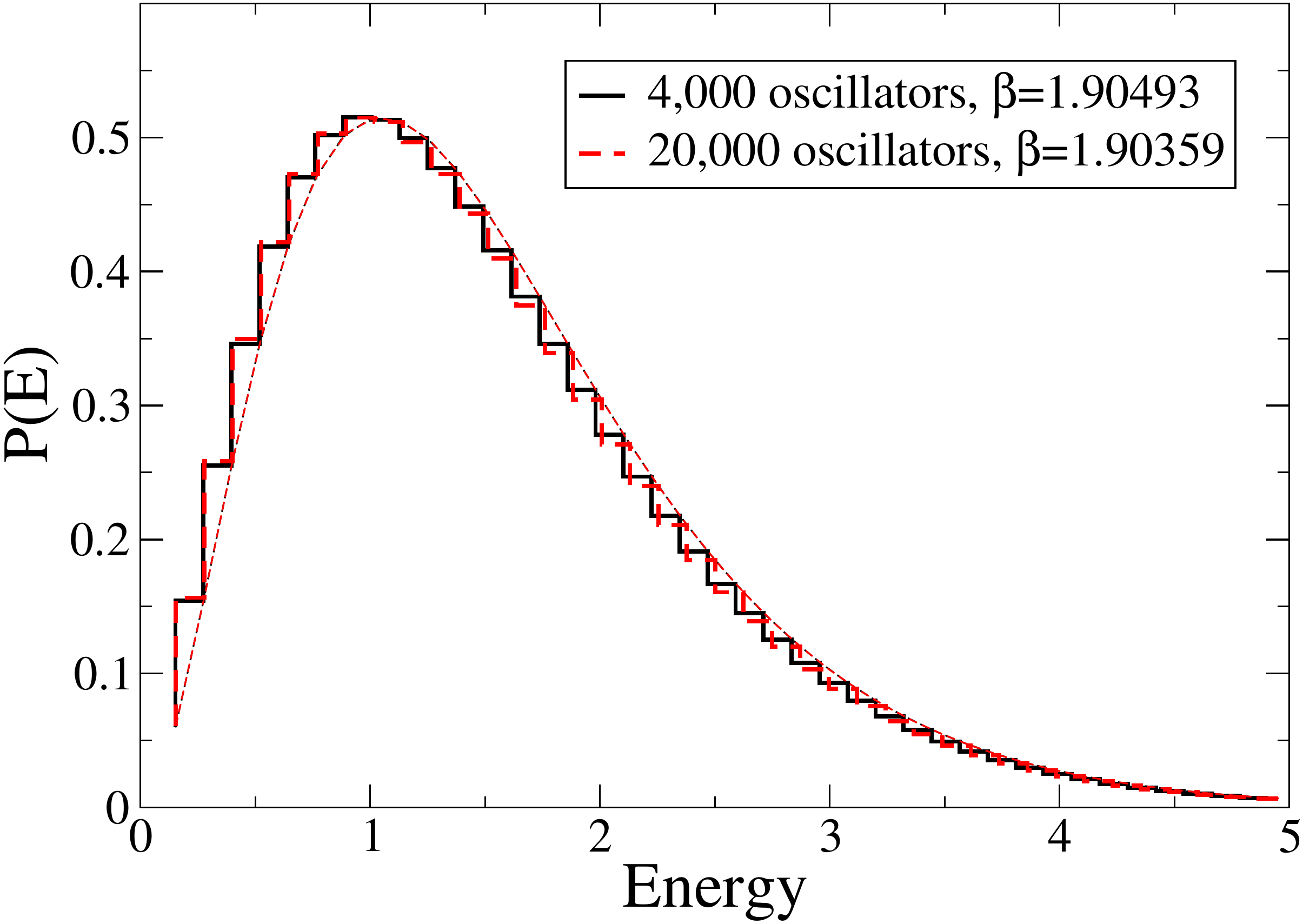}
\caption{Energy probability distributions at initial time for $4 \times 10^3$ (continuous black histogram) 
and $2 \times 10^4$ (dashed red histogram) harmonic oscillators in 3D at a nominal inverse temperature 
of $\beta=2.0$ as set up in the random number generator part of the code for assigning the initial energies 
for each atom. The inverse temperatures reported in the label arise from a single parameter fit using the probability 
distribution of Eq. (5), and the two fitting curves (dashed curves) are practically indistinguishable. 
The inverse temperature resulting from the best fit is systematically smaller by about 5 $\%$ with 
respect to the predetermined one, a discrepancy due to the coarse graining occurring in binning the distributions. 
Note that the binning is chosen to be identical in the two situations.}
\label{Fig1}
\end{figure}

In general, cold atomic gases live in higher dimensions, especially in the classical limit for 
which freezing of one or two dimensions does not occur except in the case of special, highly 
nonlinear, trapping potentials. The one-dimensional case is particularly simple, and for this 
reason somewhat misleading, as exemplified by the constant density of states in 
the limit of non-interacting harmonic oscillators. Consider the case where $N_p=1$, 
if each oscillator of the bath is described by the Hamiltonian $H=p^2/2m+m \omega^2 q^2/2$ (therefore 
neglecting the influence of the $\gamma_E$-dependent interaction term $H_{\mathrm{int}}$ as in 
Eq. (4) of  Ref.~\cite{OnSun}, in a sort of weak-coupling approximation) the phase space surface area 
of the energy shell $E$ is simply

\begin{equation}
S=\oint p dq= \pi (2mE)^{1/2} {\left(\frac{2E}{m\omega^2}\right)}^{1/2}= 2 \pi \frac{E}{\omega}\;.
\end{equation}
This implies that the number of quantum states with energy less than  or equal to $E$ is 
$S/(2\pi\hbar)=E/\hbar \omega$. The density of states -- the number of states per unit of energy 
-- will then be $g(E)={\partial S}/{\partial E}=(\hbar \omega)^{-1},$ {\it i.e.} independent of the energy. 
In general, in D-dimensions the same procedure yields a density of states $g(E)=D {E^{D-1}}/{(\hbar \omega)^D}$,
such that the energy probability distribution will be $P_D(E)={\beta^D E^{D-1} e^{-\beta E}}/{\Gamma(D)} $, 
where $\beta$ is the inverse temperature and $\Gamma$ the gamma-function. 
Specifically, for the interesting cases in 1D, 2D, and 3D, we have, respectively

\begin{eqnarray}
P_1(E)&=&\beta e^{-\beta E}, \\
P_2(E)&=&\beta^2 E e^{-\beta E}, \\ 
P_3(E)&=&\frac{1}{2} \beta^3 E^2 e^{-\beta E}.
\label{dimdist}
\end{eqnarray} 
As we shall see shortly, these relations can be used to approximately compute $\beta$ over the
transition time from non-equilibrium to equilibrium dynamics. Note that deviations from the equilibrium
distributions can also be estimated by computing $\beta$ from different combinations of higher moments
of the distribution. The consistency of these values is an indication of the appropriateness of the full
Boltzmann distribution.

In Fig.~\ref{Fig1} we show the energy probability histogram and the related best fit according to the 
expression in Eq. (5), for two initial configurations differing only in the number of atoms. 
A comparison between the inverse temperature used for the energy configuration and the value 
achieved by a best fit to the histogram provides a quantitative assessment of the accuracy of the 
obtained inverse temperature $\beta$. Figure~\ref{Fig1} illustrates that there is no significant improvement 
in the determination of the inverse temperature with a fivefold increase in the number of particles, with the
same binning used for the histograms. 
This shows, further, that the choice of binning is a greater bottleneck to accurate determinations 
of the inverse temperature than the size of the sample. From the standpoint of numerics, therefore, dealing 
with small number of particles ($N=500-1000$) is sufficient to determine $\beta$ to within a few 
percent, and within reasonable computational times.

\begin{figure*}[t]
\begin{center}
\includegraphics[width=0.45\textwidth,clip=true]{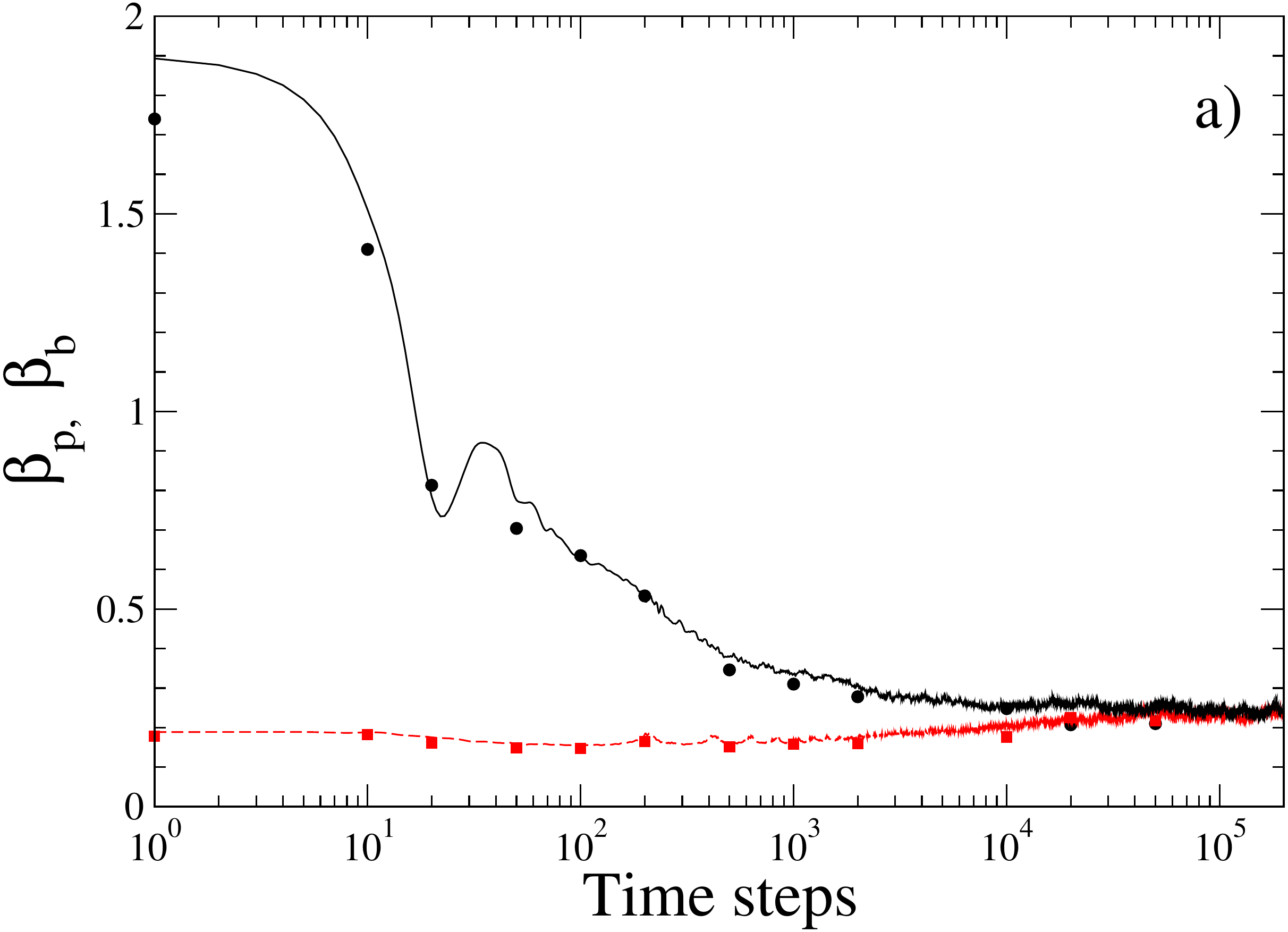} \hspace{0.1in}
\includegraphics[width=0.45\textwidth,clip=true]{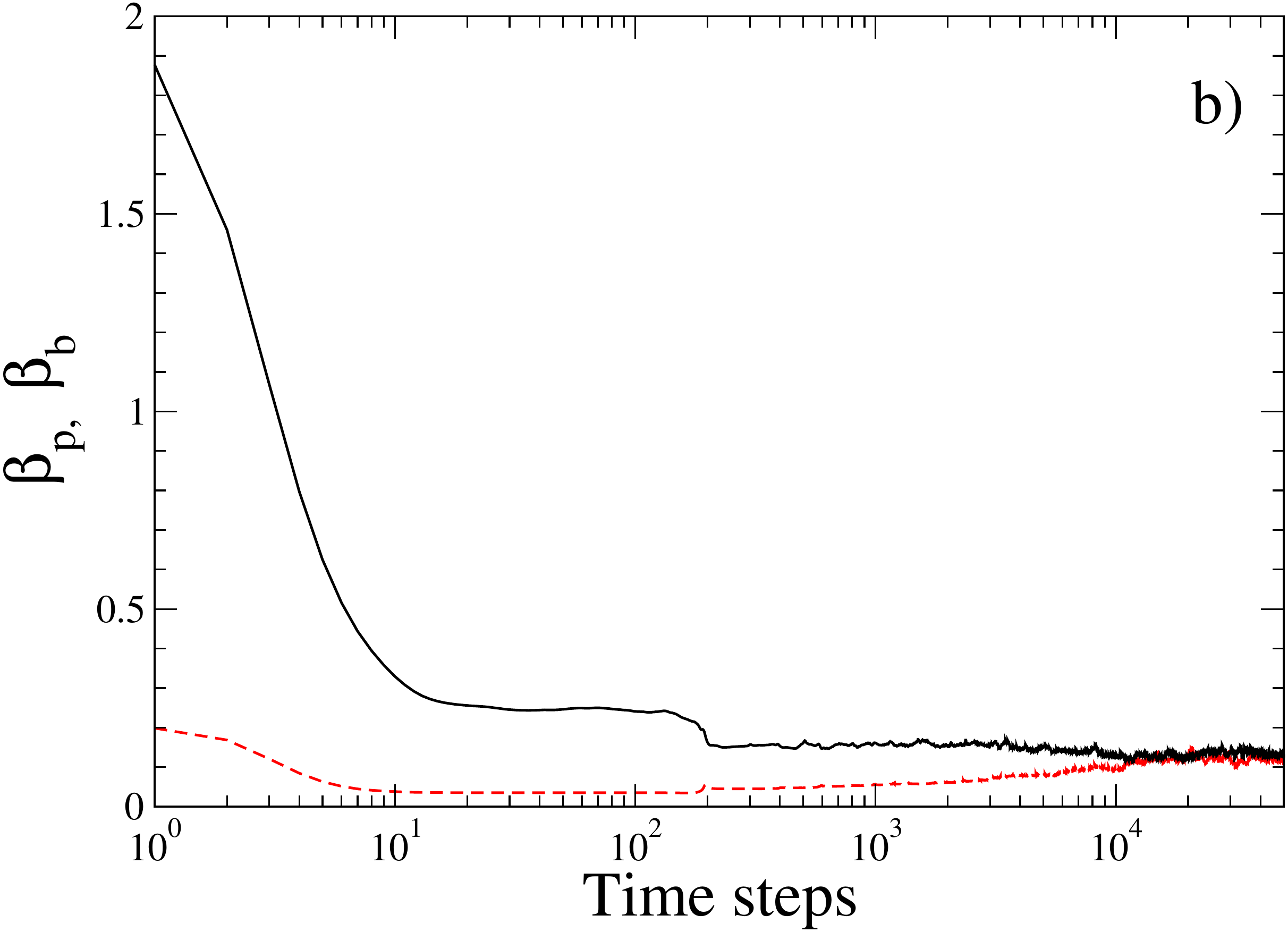}\\
\includegraphics[width=0.45\textwidth,clip=true]{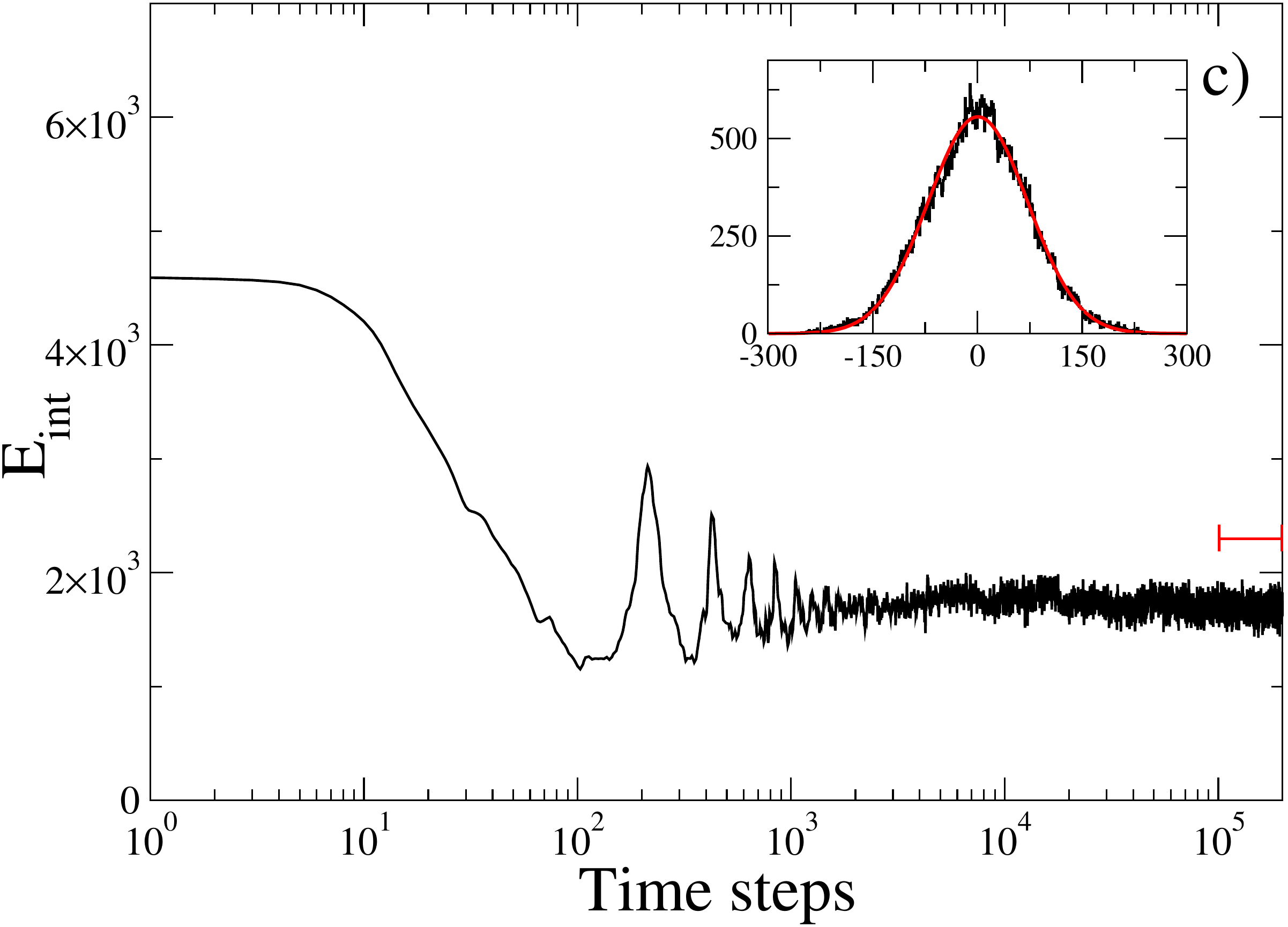} \hspace{0.1in}
\includegraphics[width=0.45\textwidth,clip=true]{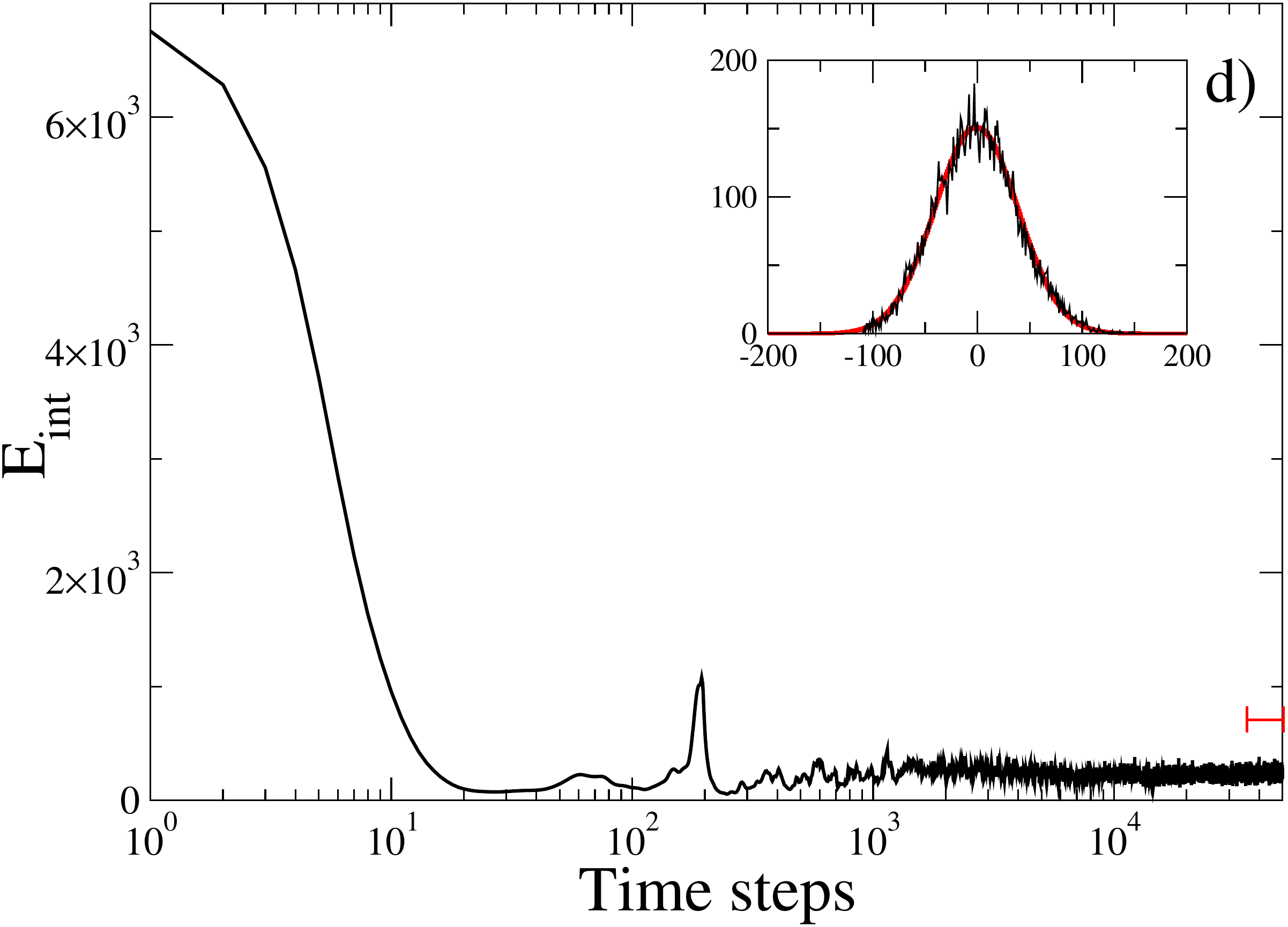}\\ 
\includegraphics[width=0.45\textwidth,clip=true]{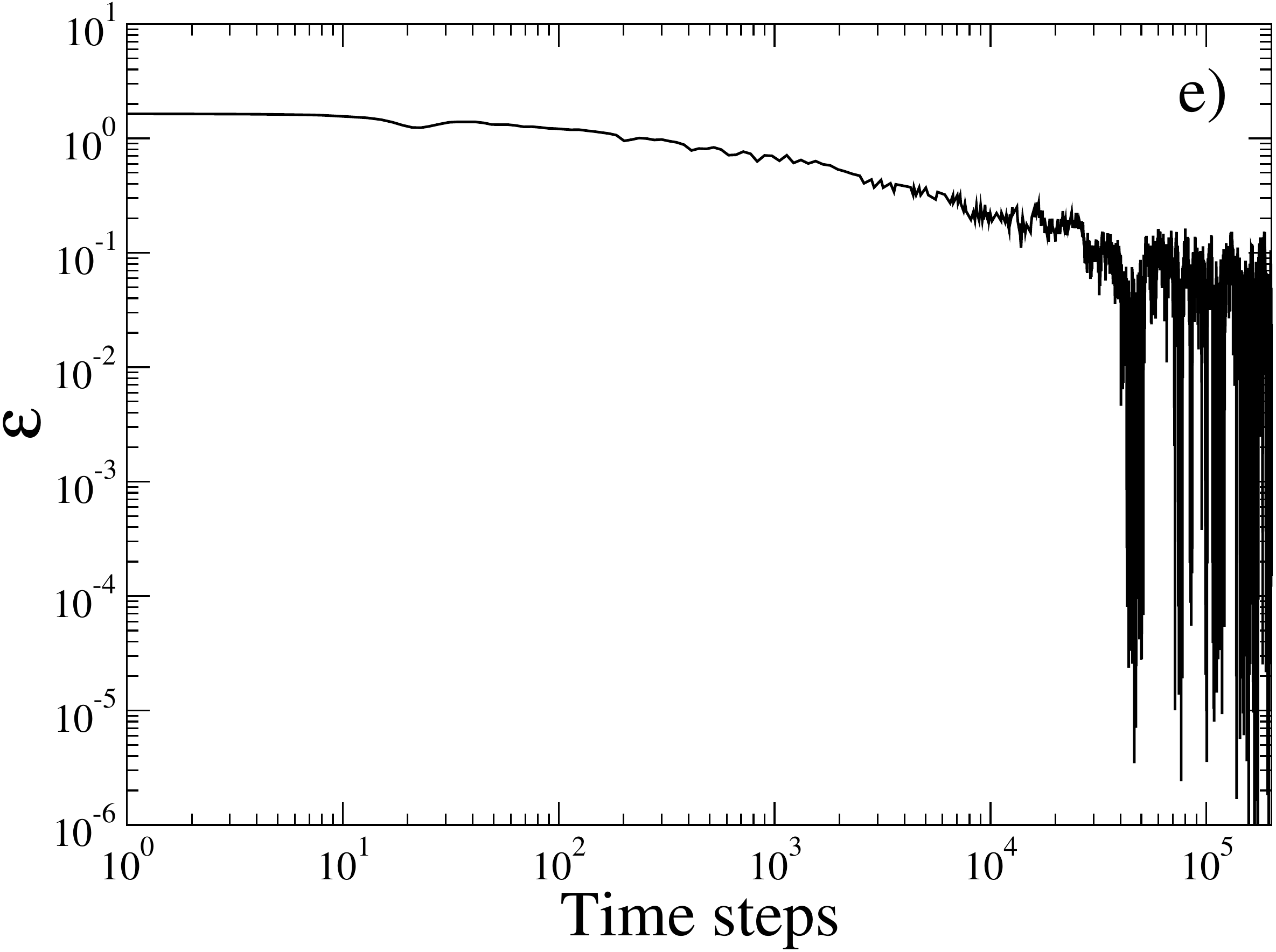} \hspace{0.1in}
\includegraphics[width=0.45\textwidth,clip=true]{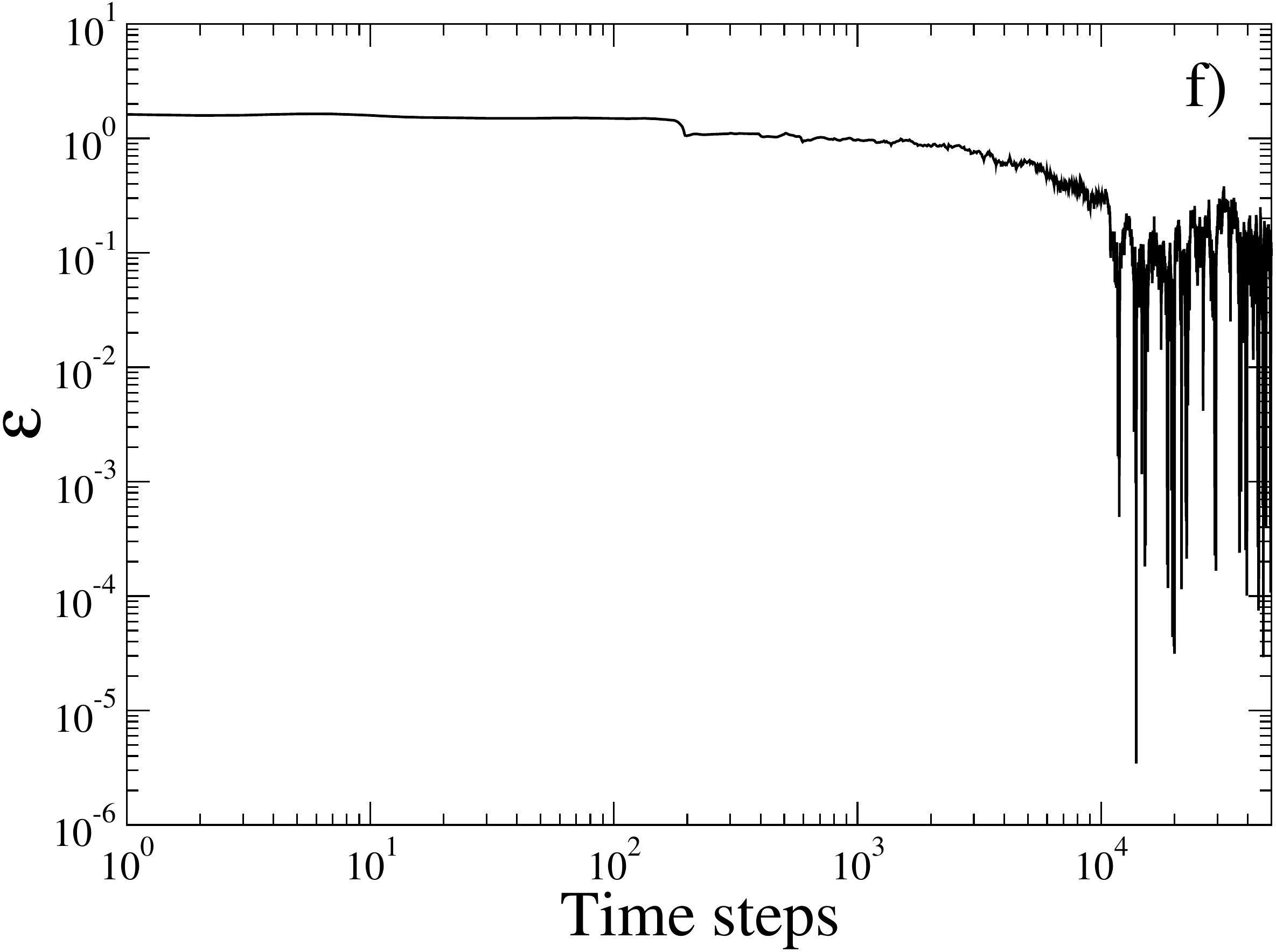}
\caption{Thermalization in one and three dimensions. (a) Evolution of inverse temperature with time 
for the two species in 1D, the cold bath particles ($\beta_b$, continuous black curve) and the warm target 
particles ($\beta_p$, dashed red curve), obtained by single run simulations. The parameters are the same 
as in Fig. 5 b in Ref.~\cite{OnSun} ($\gamma_E=0.1$, $\lambda=0.1$, $N_p=N_t=10^3$,$m=m_0=1.0$,$\omega=1.0$ and 
$\Omega=144/89$), and the dots represent the data reported in the latter paper using a fitting procedure for 
the energy distribution of the baths at specific times, while the line is the resulting curve from the 
determination of the inverse temperature through evaluation of the average energy and the average squared energy 
at each time step using  Eq.~(\ref{betaeq}).
Panel (b) shows that the same metric can be also used to track thermalization in the full 3D case, for 
$\gamma_E=14$, $\lambda=0.5$, $N_p=N_b=200$. Panels (c) and (d) show the evolution of the total 
interaction energy with time corresponding to (a) and (b), respectively. In both cases, the episodic 
interactions between the two species get increasingly random with time until it becomes a noisy, 
nearly constant value. The insets show the distribution of fluctuations around the constant value 
which is clearly well-described by Gaussians (red curves in the insets), evaluated over the intervals
indicated by the horizontal segment in the the bottom-right part of the 
main figures. Panels (e) and (f) show the time-dependence of the parameter $\epsilon$ and its effectiveness 
in quantifying the onset of thermalization.}
\label{Fig2}
\end{center}
\end{figure*}

The definitions given above allow also for a practical 'shortcut' to dynamically evaluate the 
temperature, since the following general expressions hold $\langle E \rangle = D/\beta$,
$\langle E^2 \rangle= D(D+1)/\beta^2$, which lead to the following relationship 
between the energy variance and the inverse temperature,

\begin{equation}
\sigma_E=(\langle E^2 \rangle-\langle E\rangle^2)^{1/2}=\frac{\sqrt{D}}{\beta}.
\label{betaeq}
\end{equation} 
The assessment of the inverse temperature via the simple energy probability densities as in Eq.~(\ref{betaeq})
is only valid, as mentioned above, in the weak coupling limit. This approximation allows for the neglect
of the deformation to the ellipsoids, representing the energy surfaces, due to the interaction energy. 
Ideally, one needs to construct the appropriate constant energy $E$ (including the interaction energy 
term) surface, count the states within the phase space volume as a function of $E$, and consider the 
derivative of its dependence. This procedure should then be repeated for different values of $\gamma_E$ 
(and possibly $\lambda$). Otherwise only a concept of 'effective' temperature, of limited validity in 
the strong-coupling limit, must be introduced. It should be noted that that this 'effective temperature' is  
already used in other contexts~\cite{Cugliandolo,Berthier}. In those studies, the effective $\beta$ 
serves as a proportionality constant between the rate of change of an inter-species correlation 
function and a companion response function. We will soon see an analogous relation emerge here as well.

\begin{figure}[t]
\includegraphics[width=0.95\columnwidth]{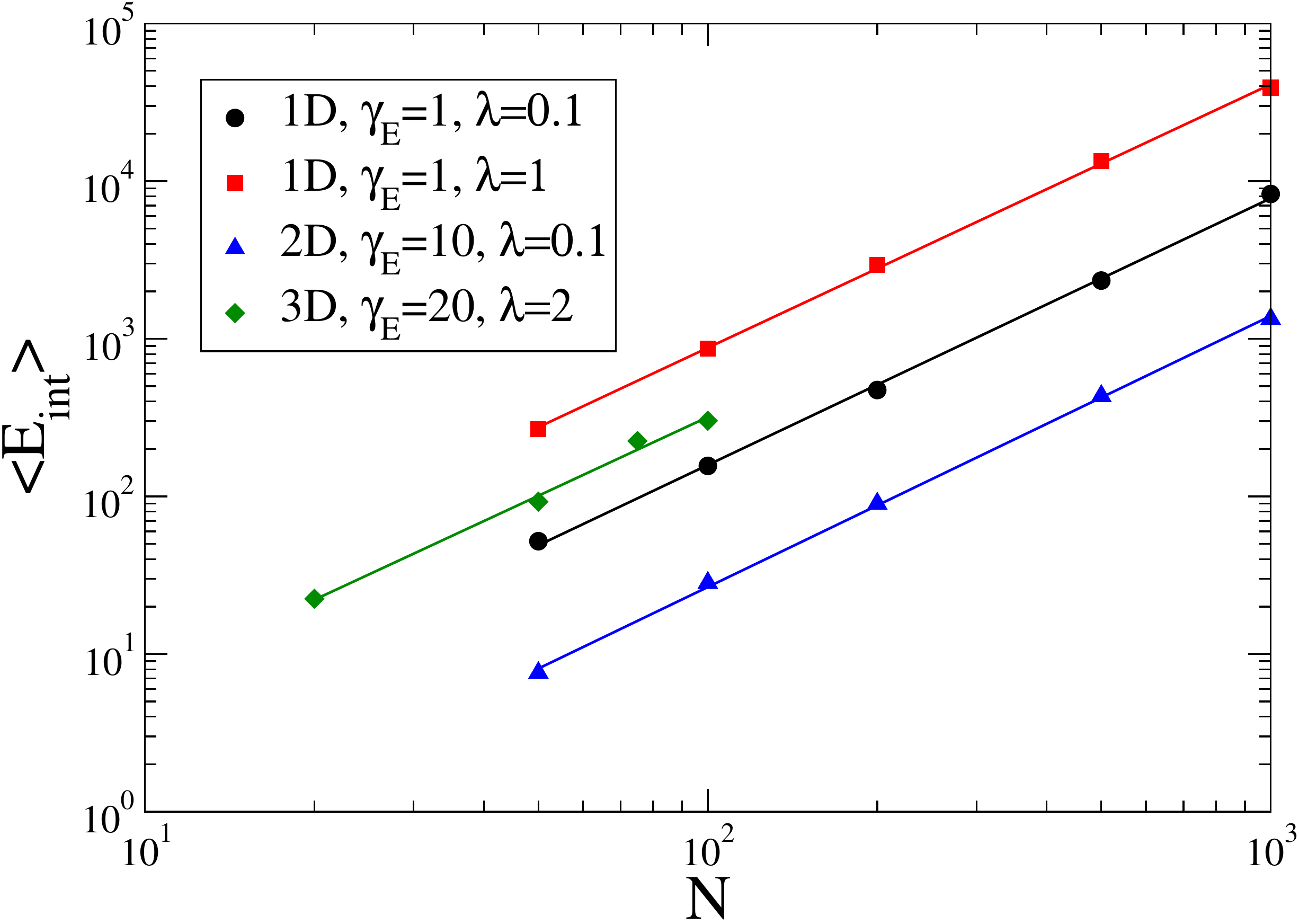}
\caption{Scaling of the long time averaged total interaction energy in 1D, 2D and 3D
versus the system size, for equal number of particles in the two systems, $N_b=N_p=N$. 
Two cases are shown in 1D corresponding to different values of the parameter $\lambda$. 
The time interval over which the average is taken is chosen based on the $\epsilon$ parameter 
with a criterion $\epsilon \leq 10^{-4}$. 
In the 1D cases, the average is taken over a longer time than in the higher dimensionality cases, 
{\it i.e.} from $80,000-100,000$ time steps. The power-law fits yield values of the exponent equal 
to $\alpha=(1.69\pm 0.03), (1.67 \pm 0.02), (1.71 \pm 0.03), (1.65 \pm 0.09)$ for the corresponding  
cases discussed in the label from top to bottom respectively. It should be noted that the suggested 
exponent of $5/3$ is expected in the large $N$ limit and when thermalization is well-established. 
In higher dimensional dynamics, the combination of smaller $N$ and incomplete thermalization results 
in the largest errors.}
\label{Fig3}
\end{figure}

Thermalization was quantified in Ref.~\cite{OnSun} by constructing the energy distribution at a set of 
fixed times and the effective $\beta$ was then determined through fitting to an exponential function. 
In order to demonstrate that constructing $\beta$ using relation Eq.~(\ref{betaeq}) works equally well to 
track the thermalization dynamics, we show in Fig.~\ref{Fig2}(a) the thermalization dynamics using Eq. (\ref{betaeq}) 
to compute $\beta(t)$ contrasted with earlier published results in~\cite{OnSun}. 
The data agree well and the clear advantage of the continuous tracking afforded by the moment 
method is seen in the additional details in the thermalization curves, in both the 1D 
and the 3D dynamics shown in Fig.~\ref{Fig2}(b). The inferences hidden in these details can be 
extracted on contrasting the relaxation dynamics with the time variation of the total interaction energy. 
This is shown in Figs.~\ref{Fig2}(c) and (d) for both 1D and 3D dynamics. The total interaction energy
is the sum of all the pairwise interactions according to the last term in Eq.~(\ref{hamil}) and the large 
excursions correspond to periodic interactions of sets of particles from the two species. 
The fact that large numbers of particles are involved is evidenced by the size of the change 
as contrasted with the value of $\gamma_E$. These episodic events are the mechanism for energy 
transfer and appear correlated with the changes in $\beta$, with the colder species being more
sensitive to these interactions. We note here that the interaction term involves pairs of particles 
each taken from one of the species. As such, it is analogous to the cross-species two-point correlation 
function whose time variation and accompanying response is used in defining effective temperatures 
(or $\beta$) in describing non-equilibrium systems. The link between the total interaction energy and 
effective temperature can be motivated by the following argument. Consider the total interaction
Hamiltonian
\begin{equation}
H_{\mathrm{int}} = \gamma_E \sum_{m=1}^{N_t} \sum_{n=1}^{N_b} \exp \left[-\frac{(q_n-Q_m)^2}{\lambda^2}\right] \;,
\end{equation}
and its ensemble average
\begin{equation}
\langle E_{\mathrm{int}} \rangle = \gamma_E N_b N_p \exp \left[-\frac{\langle (q_b-Q_p)^2\rangle}{\lambda^2}\right]\;,
\end{equation}
where the subscripts indicate a pair of particles drawn from the bath ($b$) and test particle ($p$) species and
the approximation $<f(x)>=f(<x>)$ is used.
We can identify the ensemble average in the exponent as the interspecies two-point correlation function which
can be now be written as
\begin{equation}
C_{bp} = \langle (q_b-Q_p)^2\rangle = 
-\lambda^2 \ln{\left( \frac{\langle E_{\mathrm{int}} \rangle}{\gamma_E N_b N_p} \right)} \;.
\end{equation}
We note here that the correlation between pairs of particles from the two species is a consequence of
interaction, therefore changes in the interaction energy over time will result in changes in the 
correlation function. This may be expressed as
\begin{equation}
\frac{dC_{bp}}{dt} = -\frac{\lambda^2}{\langle E_{int} \rangle} \frac{d\langle E_{\mathrm{int}}\rangle}{dt} \;.
\end{equation}
Contrasting this expression with definitions used for effective temperatures suggests that the  
total interaction energy $\langle E_{\mathrm{int}}\rangle$ here plays the role of $k_B T_{\mathrm{eff}}$ 
in~\cite{Cugliandolo,Berthier,Speck}. This relationship also makes clear that stationarity of the correlation 
function results in the stationarity of the total interaction energy and, hence, the effective temperature.

\begin{figure*}[t]
\begin{center}
\includegraphics[width=0.31\textwidth,clip=true]{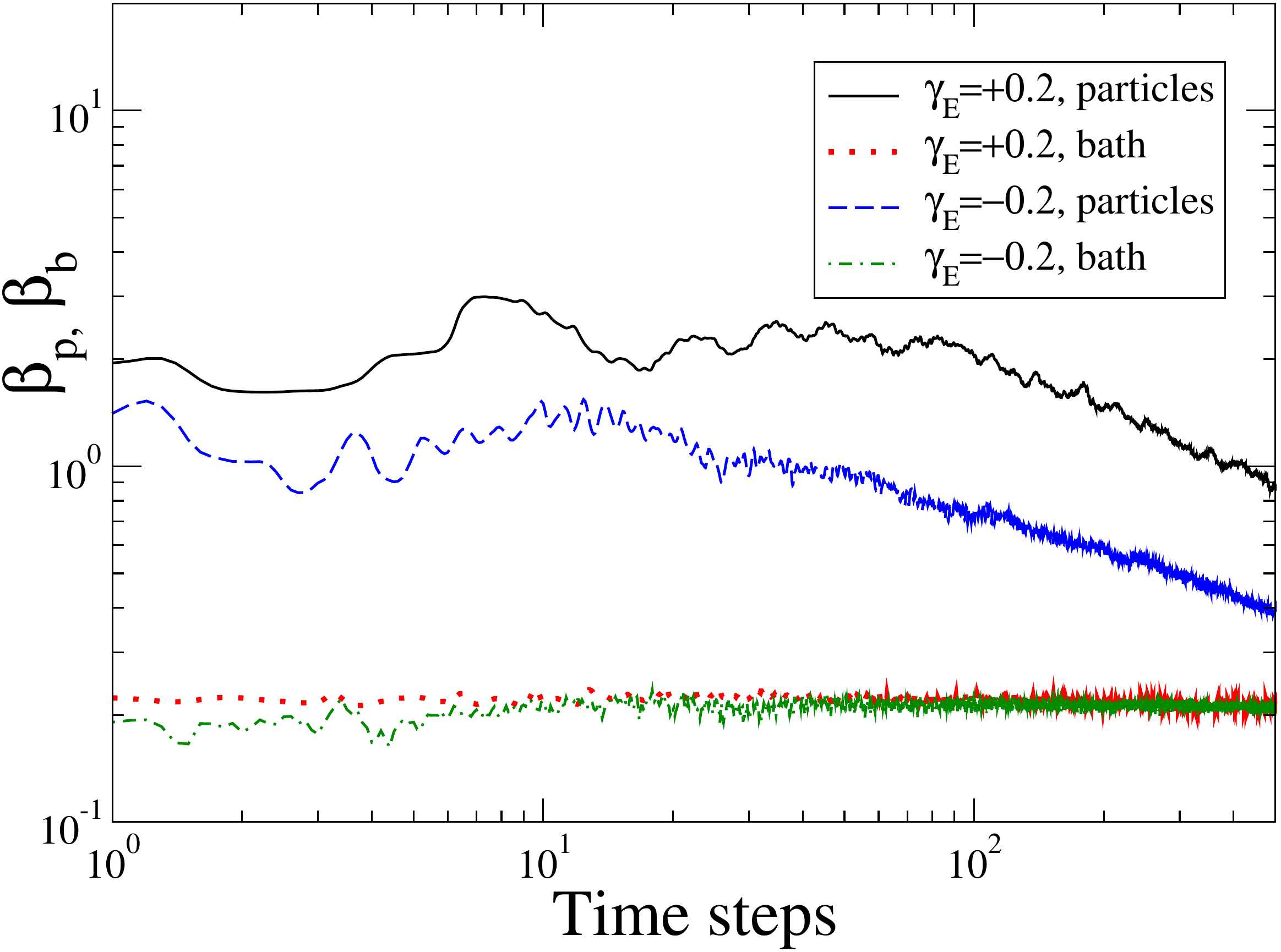}
\includegraphics[width=0.31\textwidth,clip=true]{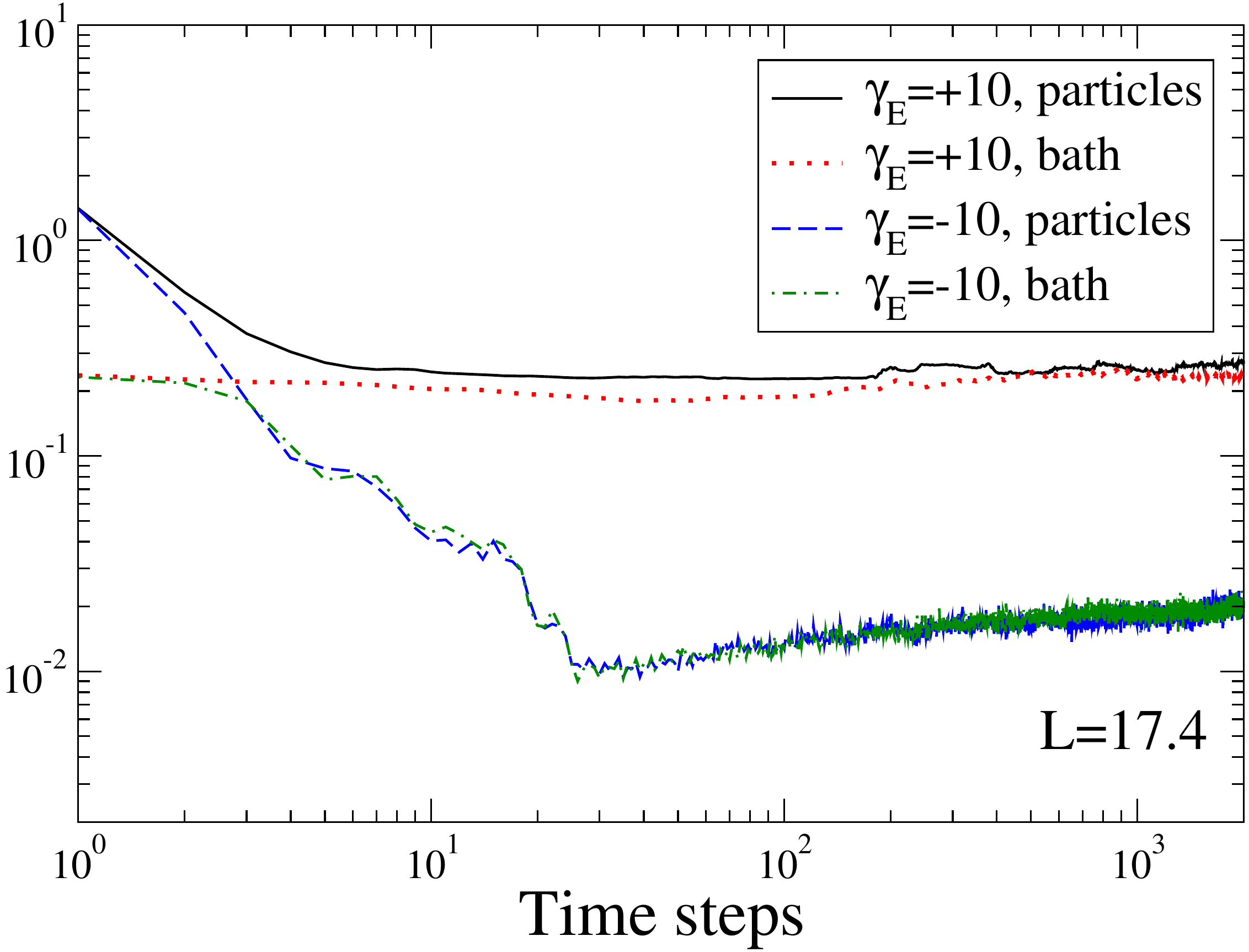}
\includegraphics[width=0.31\textwidth,clip=true]{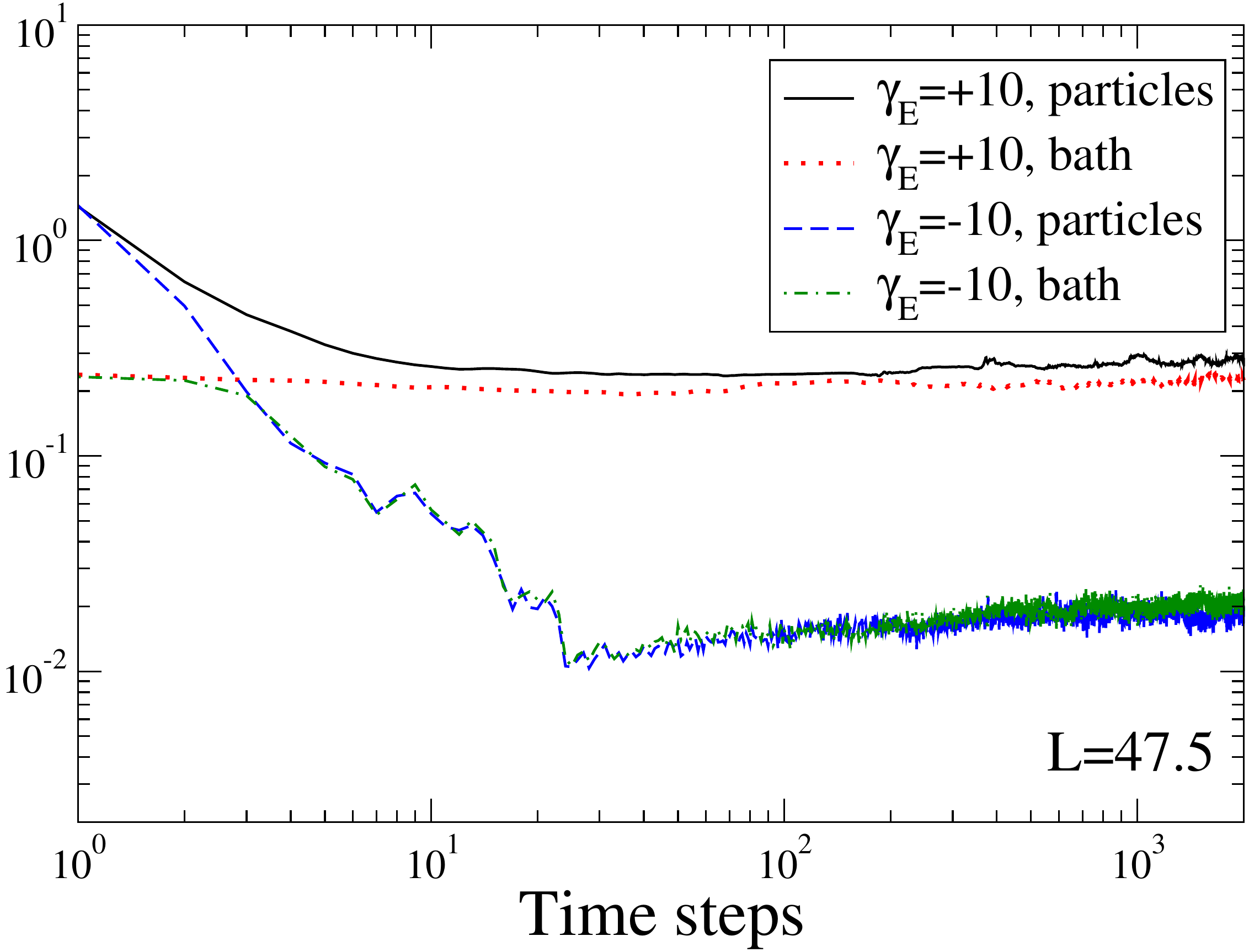}
\caption{Comparison between the dynamics of thermalization in (a) 1D where the only difference 
is in the sign of the coupling strength $\gamma_E$. (b) and (c) are the analogous effects explored 
in 3D where the difference between the two plots is the total angular momentum of the two-species 
system which affects the number of inter-species encounters, as indicated in the lower right corner 
of each plot.}
\label{Fig4}
\end{center}
\end{figure*}

Before moving to a more detailed discussion of the inferences seen in the behaviour of the interaction 
energy with time, we draw attention to Figs.~\ref{Fig2}(e) and (f) which show a simple but seemingly 
effective way of quantifying the thermalization time. What is shown is the quantity
\begin{equation}
\epsilon = 2 \frac{|\beta_p-\beta_b|}{\beta_p+\beta_b} \;,
\end{equation}
which relates the absolute difference in $\beta$ between the two species with their mean value. 
Contrasting the thermalization plots in Fig.~\ref{Fig2}(a) and (b) with the $\epsilon$ behaviour 
clearly suggests the possible use of this measure to numerically quantify thermalization times. 
Notice that when the $\epsilon$ parameter drops below a given threshold, say $\epsilon \leq 10^{-3}-10^{-4}$, 
the number of spikes with values lower than the threshold increases significantly even if occasionally 
the parameter raises again to higher values. These fluctuations in the thermalization dynamics are more
clearly visible in $\epsilon$ than in the inverse temperature curves. The spikes are correlated to the details of the
thermalization dynamics, and in the absence of a damping mechanism can last for long times. 
Their duration is inversely proportional to the instantaneous rates when the two $\beta$ curves cross each other. 
Inverse temperatures which are very close to each other for a large extent of time therefore results in a broad 
time interval below threshold. Additionally, we expect the finite number of particles to also play a role with 
$\epsilon$ being a more effective metric with increasing $N_b, N_p$, one which could be useful in quantitatively 
contrasting cooling strategies.

Returning to the interaction energy, a noisy plateau is seen for longer times coinciding with the thermalization 
of the two species. In this time domain, the time-averaged mean value 
can be computed and the distribution of fluctuations about this mean constructed. This is shown in the 
insets in Figs.~\ref{Fig2}(c) and (d) and both are clearly Gaussian. Thus the total energy (Hamiltonian) 
which was originally distributed as $H_p+ H_b+ H_{int}$ evolves with time to the approximation 
$H'_p+ H'_b+ G(t) + {\mathrm{const}}$ where $G(t)$ represents the Gaussian fluctuations in the interaction
energy. As the interaction energy was connected to an effective $\beta$, these can be viewed as akin to thermal
fluctuations. 
This is also an indicator of the onset of thermalization and we recover the textbook, non- or weakly 
interacting construct used when defining macro variables such as temperature. 
As seen from Fig.~\ref{Fig2}, this behaviour is seen irrespective of the dimensionality of the dynamics 
suggesting a universal mechanism for thermalization. The energy content in the mean (constant) value, 
as contrasted with that given by the initial temperatures $T_p$ and $T_b$ of the two subsystems, 
determines whether or not the final equilibrium temperature is hotter than either of two initial 
ones, a point which we will return to later. 
We note here that, as a consequence, varying the strength of the inter-species interaction can be used 
as a control parameter to achieve specified final states~\cite{Rahmani}.
We anticipate that the exact functional form of the interaction only affects the 
timescales for thermalization rather than the universal behaviour itself. The expectation is based on the 
time evolution of the interaction energy which transitions from deterministic 
interactions between the two species, based clearly on the details of the dynamics, to a probabilistic 
regime described by an average interaction per particle with small fluctuations. This suggests homogenization
of the neighbourhood around each test particle, meaning that the two atomic species are well-mixed.
This is a view which is consistent with another result seen in the numerics, suggestive of a possible connection
to a result from fluid dynamics.

\begin{figure}[b]
\begin{center}
\includegraphics[width=0.95\columnwidth, clip=true]{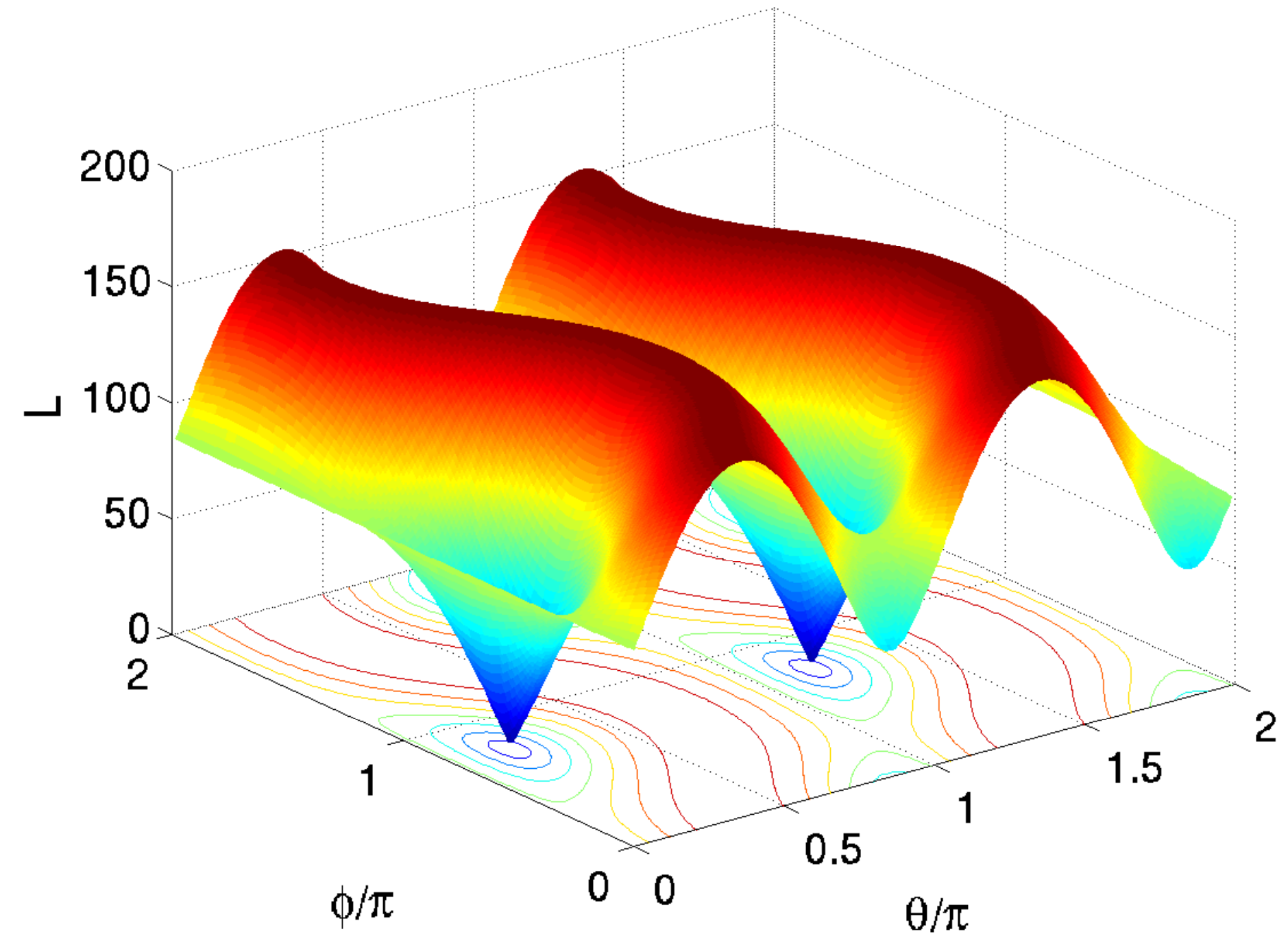}
\caption{Total angular momentum of the atomic clouds versus the angles $\theta$ 
and $\phi$ between the momentum and the position of each particle for given inverse 
temperatures $\beta_p=0.2$ and $\beta_b=2.0$. The behaviour is identical when considering
the differential angular momentum of the atomic clouds. The profile shown does depend on
the particle numbers in each species and here we use $N_b=N_p=500$.}
\label{Fig5}
\end{center}
\end{figure}

We consider the time-averaged total interaction energy as a function of the system size 
(number of oscillators) where $N_b=N_p=N$. This was done both for one- and higher dimensional dynamics. 
In each case, the parameter values considered resulted in thermalization over the total time scale 
considered. We note that for higher dimensions, a larger interaction strength was considered to ensure 
thermalization over a shorter timescale, simply for numerical considerations. The results, as 
seen from Fig.~\ref{Fig3}, display power-law scaling with system size with a common exponent 
(within fitting errors) independent of dimensionality. In all instances, the time averaged 
interaction energy scales as $N^{\alpha}$ where $\alpha$ is close to $5/3$, an exponent which has
significance in another context.

This value of the scaling exponent is reminiscent of the Kolmogorov scaling , as $k^{-5/3}$ where 
$k$ is the wavenumber, associated with turbulent mixing~\cite{Kolmogorov1,Kolmogorov2}. 
In that case, this value can be recovered using elegant, dimensional arguments involving viscosity 
and an energy transfer rate. In our situation, the inter-species interaction is responsible for 
both energy transfer as well as an 'effective damping' mechanism. Nevertheless, we speculate that 
the mechanism for eliminating thermal heterogeneity is analogous to the removal of density or thermal 
gradients in the fluid problem. Here, we move from a deterministic description to a statistical one 
as evidenced by the behaviour of the interaction energy. Notice that, while the Caldeira-Leggett model 
allows for thermalization, due to the linear nature of the interatomic interactions there is no energy 
transfer between different timescales, despite the presence of a continuum of frequencies 
in the thermal bath. Here instead we expect mixing among all timescales due to 
the nonlinear interaction, in spite of the monochromatic baths. This occurs in 
a manner similar to the mixing of different lengthscales in the turbulent mixing of fluids. 
A more detailed exploration of this intriguing analogy is ongoing. In the current context, we consider the 
scaling result another quantitative indicator of universal aspects of the thermalization 
mechanism.

\begin{figure}[t]
\begin{center}
\includegraphics[width=0.90\columnwidth, clip=true]{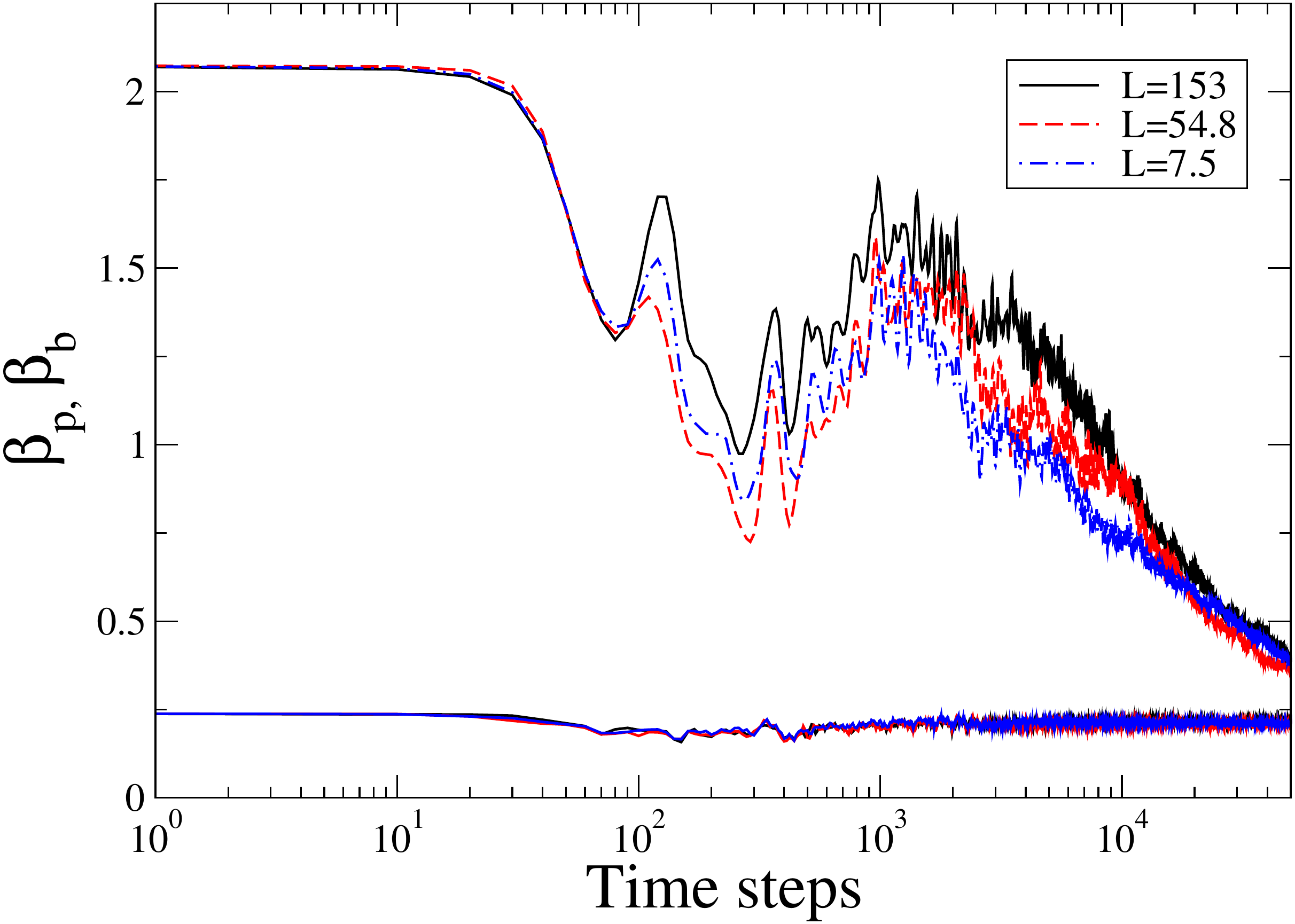}
\caption{Comparison between thermalization trajectories for initial conditions differing 
only by the amount of total angular momentum, with $N_b=N_p=500$, $\gamma_E=-0.1$ and $\lambda=1.0$ in all
cases. The total angular momentum values are indicated in the legend. The low value of $\gamma_E$ has
been chosen to emphasize differences due to total angular momentum.}
\label{Fig6}
\end{center}
\end{figure}

\section{Singular Aspects of Higher-Dimensional Thermalization}

It is clear that the dynamics of thermalization in higher dimensions depends on aspects that
are not present in the simpler 1D case. Significant amongst these is angular momentum which brings
with it the ability to avoid interactions, notably head-on collisions, thus inhibiting energy
exchange which, in turn, slows down the thermalization process. Of course, changing the nature
of the interaction from repulsive to attractive brings in aggregation mechanisms which force
interactions. 

We begin by studying the dependence of the dynamics on the sign of the interaction potential. 
As shown in Fig.~\ref{Fig4}(a), the use of attractive interactions results in a faster thermalization 
in 1D with respect to the repulsive interaction with the same absolute value of $\gamma_E$. 
This can be easily interpreted from the propensity for the two clouds to cluster in the former case.  
In an experimental scenario, however, this effect will be accompanied by larger three-body decay rates 
for attractive interactions also due to the higher densities, with consequent increases in heating rates. 
The simulations in 3D , for the same range of parameters used in 1D, show that thermalization in general 
seems to occur, if at all, on considerably longer timescales. This may be interpreted in configuration space 
as due to the presence of a multitude of 'avoiding' trajectories, and of a partial overlap between the 
two atomic clouds. Of course, increasing the interaction strength $\gamma_E$ or the range $\lambda$ or the 
sign of the interaction are all ways to speed up the process. Figure~\ref{Fig4}(b) shows the
role of larger $|\gamma_E|$ in achieving this. 

\begin{figure}[t]
\begin{center}
\includegraphics[width=0.95\columnwidth, clip=true]{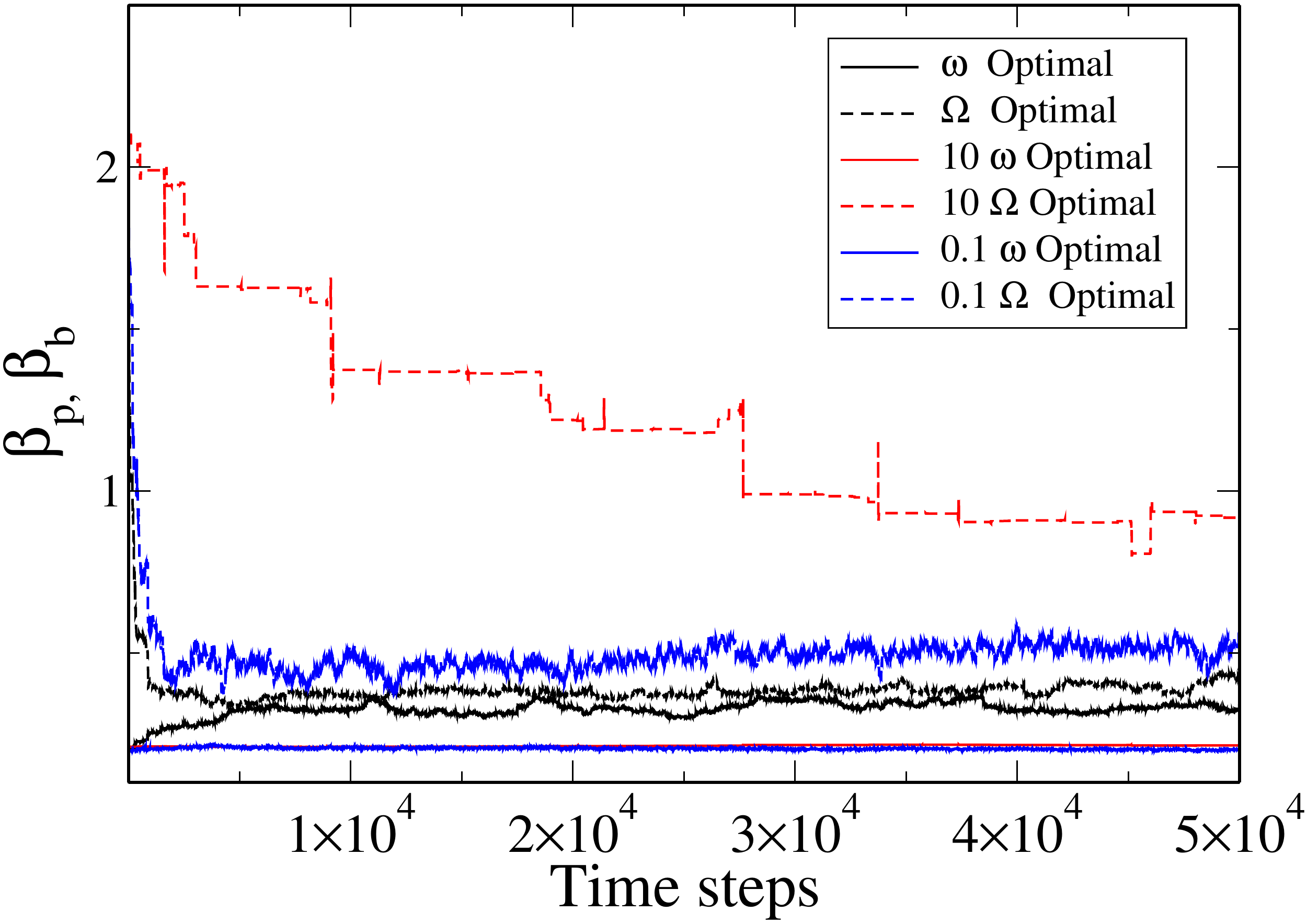}
\caption{The effect of changing the frequency ratio $\Omega/\omega$. 
As indicated in the legend, the optimal ratio suggested in the text and significantly 
smaller and larger values are shown. In all cases, the other parameters are 
$T_b=0.5, T_p=5.0, N_p=N_b=200, \gamma_E=14$ and $\lambda=0.1$.}
\label{Fig7}
\end{center}
\end{figure}

Additionally, as the particles also possess angular momentum, they are not forced as in the 1D  
case to have $s$-wave-like encounters with the other particles. Therefore, all other conditions for the atomic 
clouds being the same, we expect slower thermalization in higher dimensions. Angular momentum is a 
meaningful quantity to be considered in this setting, as both the harmonic trapping potential and 
the interspecies interaction are central forces, therefore the total angular momentum available in 
the mixture is conserved. In principle, the Boltzmann distribution of each cloud depends on all 
independently conserved quantities, including angular momentum (see supplementary information in 
~\cite{Guery} for a detailed discussion), but thermalization leads to a stationary state in which 
its role is expected to be marginal. In order to quantify these hypotheses we have 
injected a controllable total angular momentum to the initial conditions of the two clouds, to see 
how the thermalization timescale depends on it. The initialization of the two baths must be rearranged 
in such a way that we control the amount of angular momentum energy. The Hamiltonian for a single particle
in spherical coordinates is $ H=m\dot{r}^2/2 + m \omega^2 r^2/2 + L^2/(2mr^2),$ where the first two terms 
represent the radial kinetic energy and the harmonic potential energy, respectively, and the third contains 
the kinetic energy due to the presence of angular momentum. This is implemented by first picking an energy 
value for each particle consistent with the Boltzmann distribution and random positions from
which the magnitude of the total momentum can be computed. 
The choice of the direction of momentum with respect to the vector $\vec{\bf r}$ can then be
controlled by means of two angles, $\theta$ and $\phi$, such that
$p_x=p \sin \theta \cos \phi, p_y= p \sin \theta \sin \phi, p_z= p \cos \theta$. 
In Fig.~\ref{Fig5}, we show the dependence of the total angular momentum of the two clouds as a function of
$\theta$ and $\phi$, chosen to have the same value for each atom. The initial
temperatures of both clouds are specified in the caption. The angular momentum differential between
the two clouds shows identical dependence on the two angles. 

The control of the angular momentum of the two clouds allows us to compare thermalization in various 
cases. We return to panels (b) and (c) of Fig.~\ref{Fig4} which is a situation where the interaction 
strength is strong. The  only difference between the two panels is the total angular momentum of the 
two clouds. Here, we expect the interaction strength to dominate and the role of the angular momentum 
is barely visible. In order to isolate any difference arising strictly from changes in the total angular 
momentum, we reduce the value of $\gamma_E$. These results are shown in Fig.~\ref{Fig6} where we compare 
three cases, minimal and maximal angular momentum, and an intermediate value. Any sensitivity to angular 
momentum is apparent only at an intermediate stage, starting after about $10^2$ time steps, and 
lasting about two decades. In the region around $10^3$ time steps the three curves of the lower 
temperature bath, in spite of statistical fluctuations, settle themselves in a systematic order approaching 
equilibrium with the hotter bath according to the content of angular momentum, and with a thermalization speed 
inversely proportional to the latter. 
Beyond about $2 \times 10^4$ time steps the difference among the curves is negligible. This is not 
too surprising given the earlier discussion on the universal route to thermalization. 
Changes in total angular momentum should only affect the deterministic phase of the thermalization 
process which, given the many-particle and nonlinear nature of the interaction, can be quite short. 
Experimentally, control of the angular momentum of a Bose-Einstein condensate has been demonstrated 
via coherent transfer from optical fields using a stimulated Raman process~\cite{Wright}, providing a 
protocol for testing these findings. As commented on earlier, we only analyse the case of velocity-independent 
interactions throughout this contribution, unlike the analysis reported in~\cite{OnSun}. 
We expect that further differences in thermalization dynamics could arise by using velocity-dependent 
interactions due to the peculiar role of the velocity distribution in assigning angular momenta.

Long-time simulations show that the approach to complete equilibrium is rather slow even for moderate values 
of interaction strength. A more sensitive (and controllable) knob which can be used to speed up thermalization 
is available by optimizing the spatial overlap between the two clouds. The unavoidable presence of atoms 
with large angular momentum and high energy in the Boltzmann tail will make these atoms very weakly 
interacting with the rest of the cloud of the other species. Therefore, no matter of how long one 
waits for thermalization, these atoms will form a sort of 'closed subsystem' at nearly 
constant energy, apart from occasional close encounters in which large amounts of energy 
will be exchanged. This can be mitigated if the overlap between the clouds is maximized at all times.
The average squared position of the atoms of each atomic species with respect to the minimum of the 
trapping potential is written in terms of the corresponding temperature as 
$\langle q^2 \rangle= 2 k_B T_b/m \omega^2$ and $\langle Q^2 \rangle= 2 k_B T_p/M \Omega^2.$
It is therefore possible to equate the mean squared positions for the two clouds if 
\begin{equation}
\frac{\Omega}{\omega}= \left(\frac{m}{M} \frac{T_p}{T_b}\right)^{1/2}.
\end{equation}
In order to have optimal overlap, one should continuously change the trapping frequency ratio in 
accordance with this expression. This possibility, discussed in~\cite{BrownPre} (in particular see Fig. 6 there), 
restates the problem in the context of a control theory approach. 

\begin{figure*}
\begin{center}
\centering 
\centering
\includegraphics[width=0.42\textwidth,clip=true]{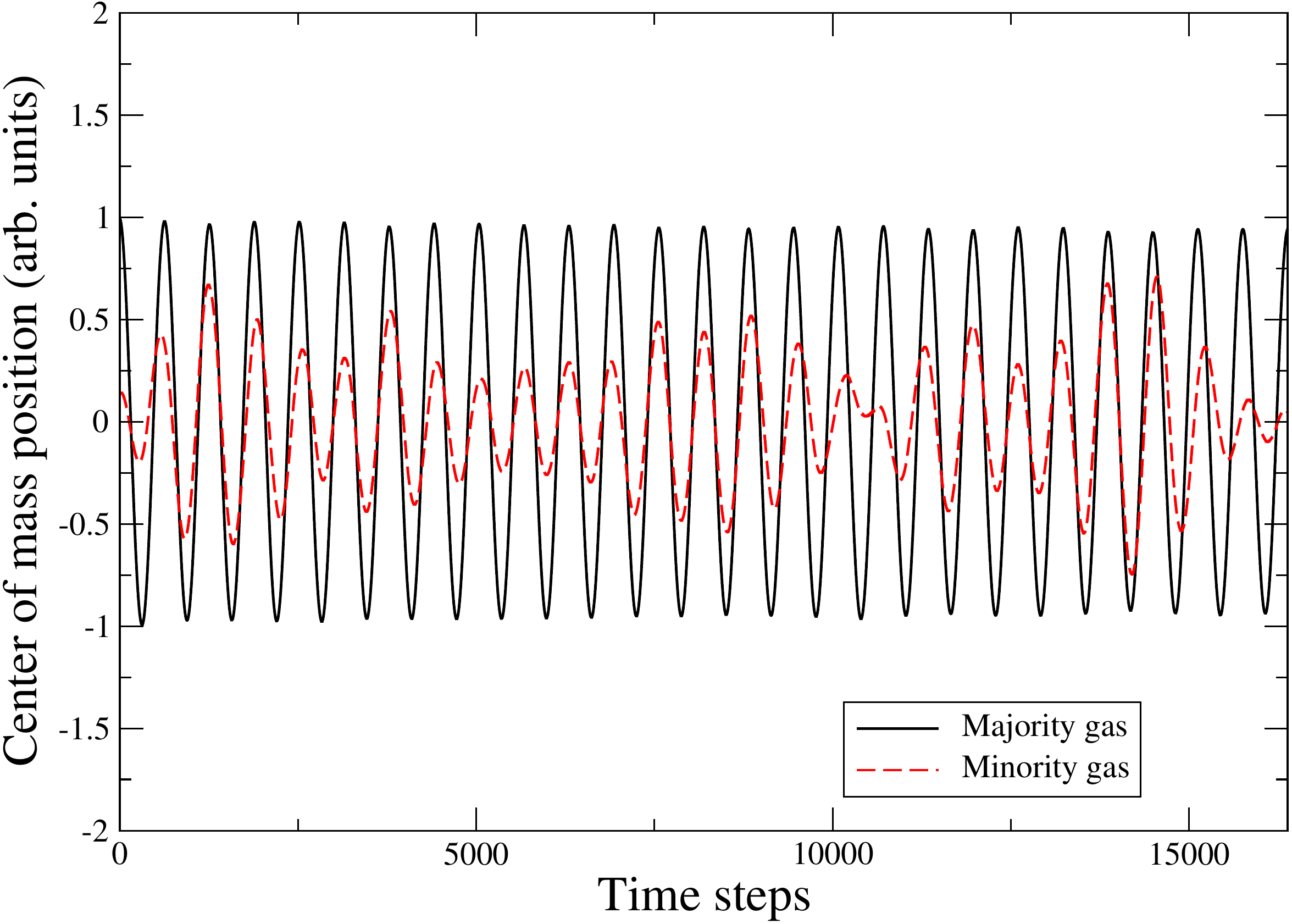} \hspace{0.1in}
\includegraphics[width=0.42\textwidth,clip=true]{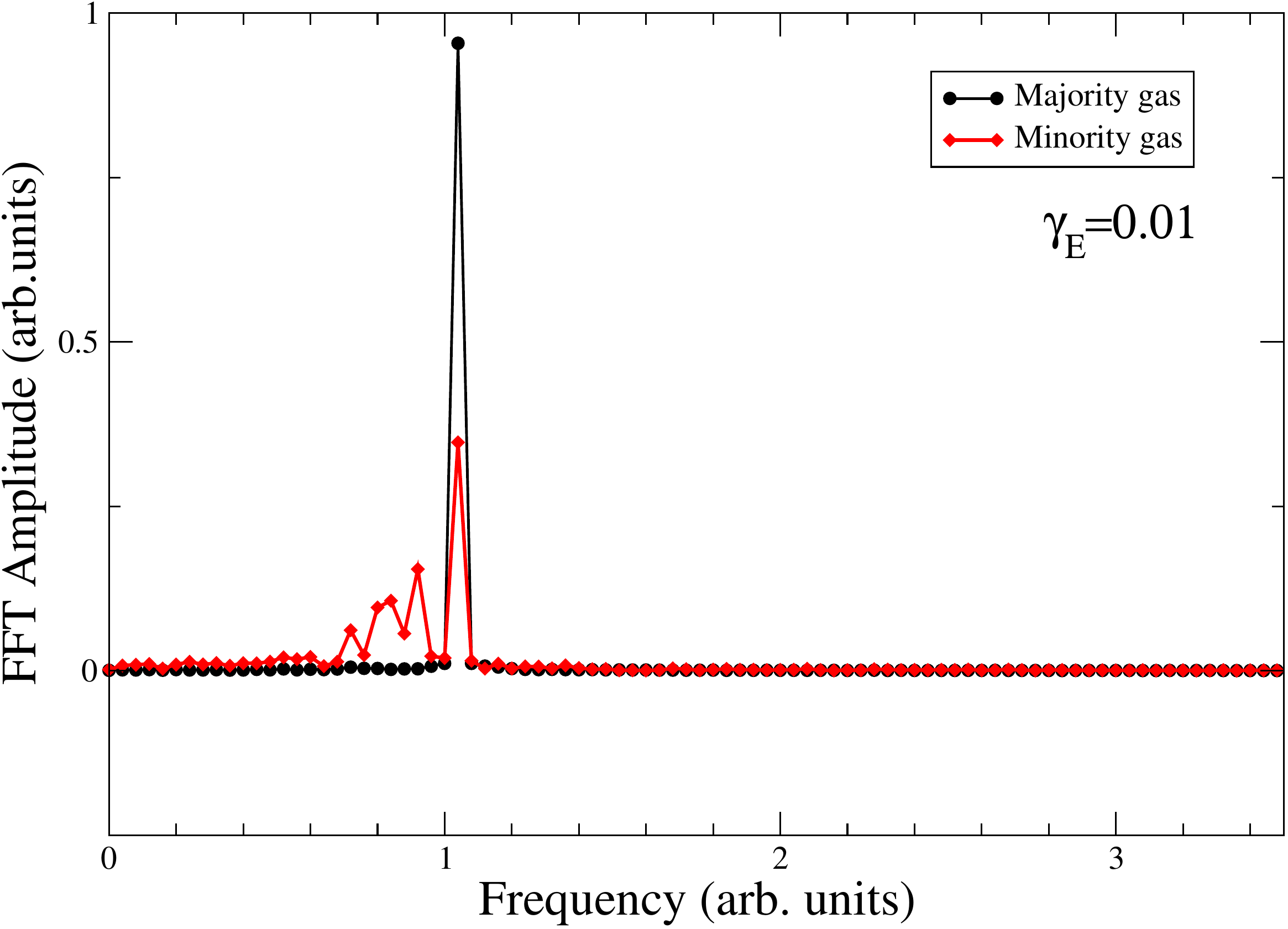} \\
\includegraphics[width=0.42\textwidth,clip=true]{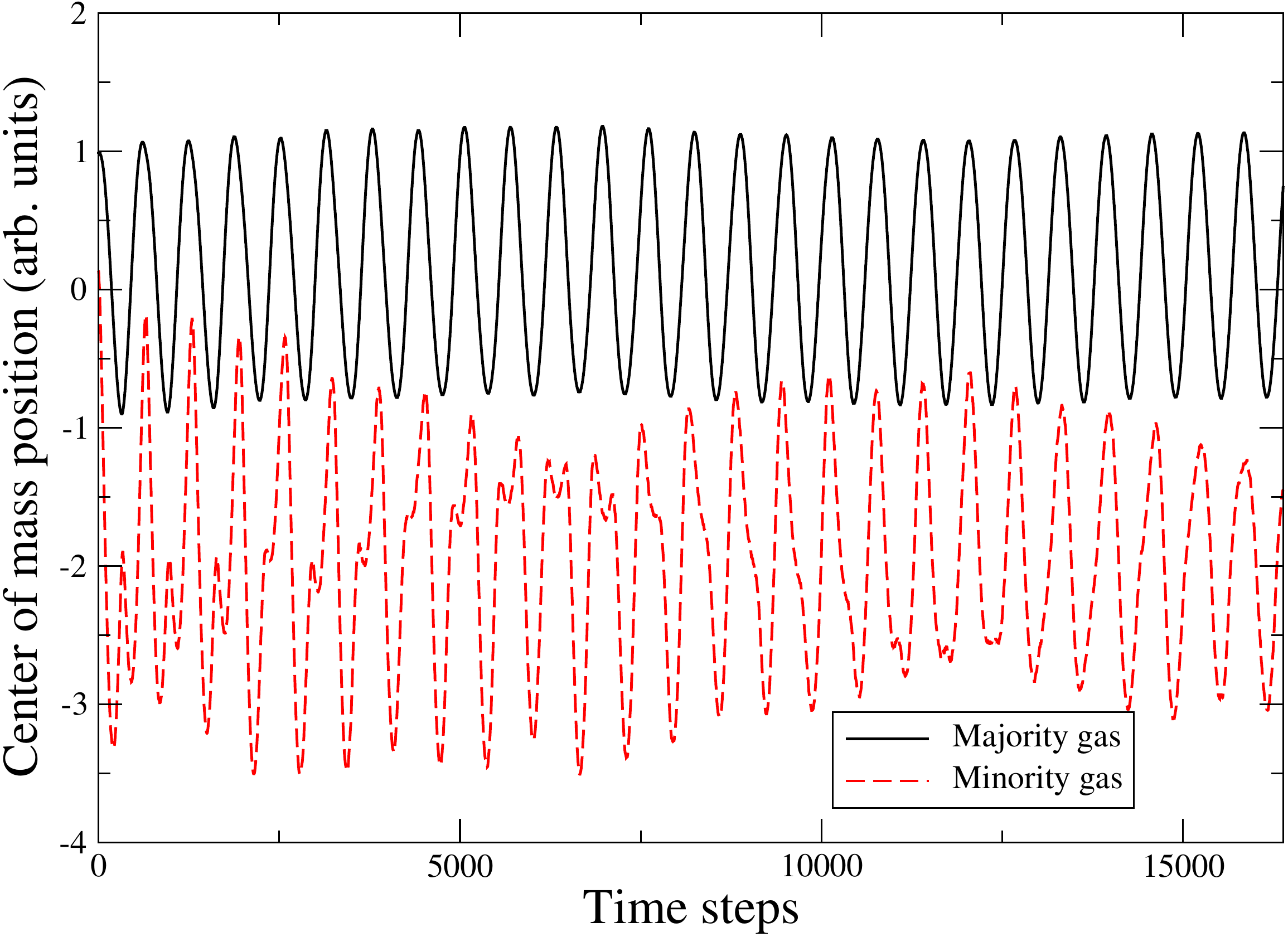} \hspace{0.1in}
\includegraphics[width=0.42\textwidth,clip=true]{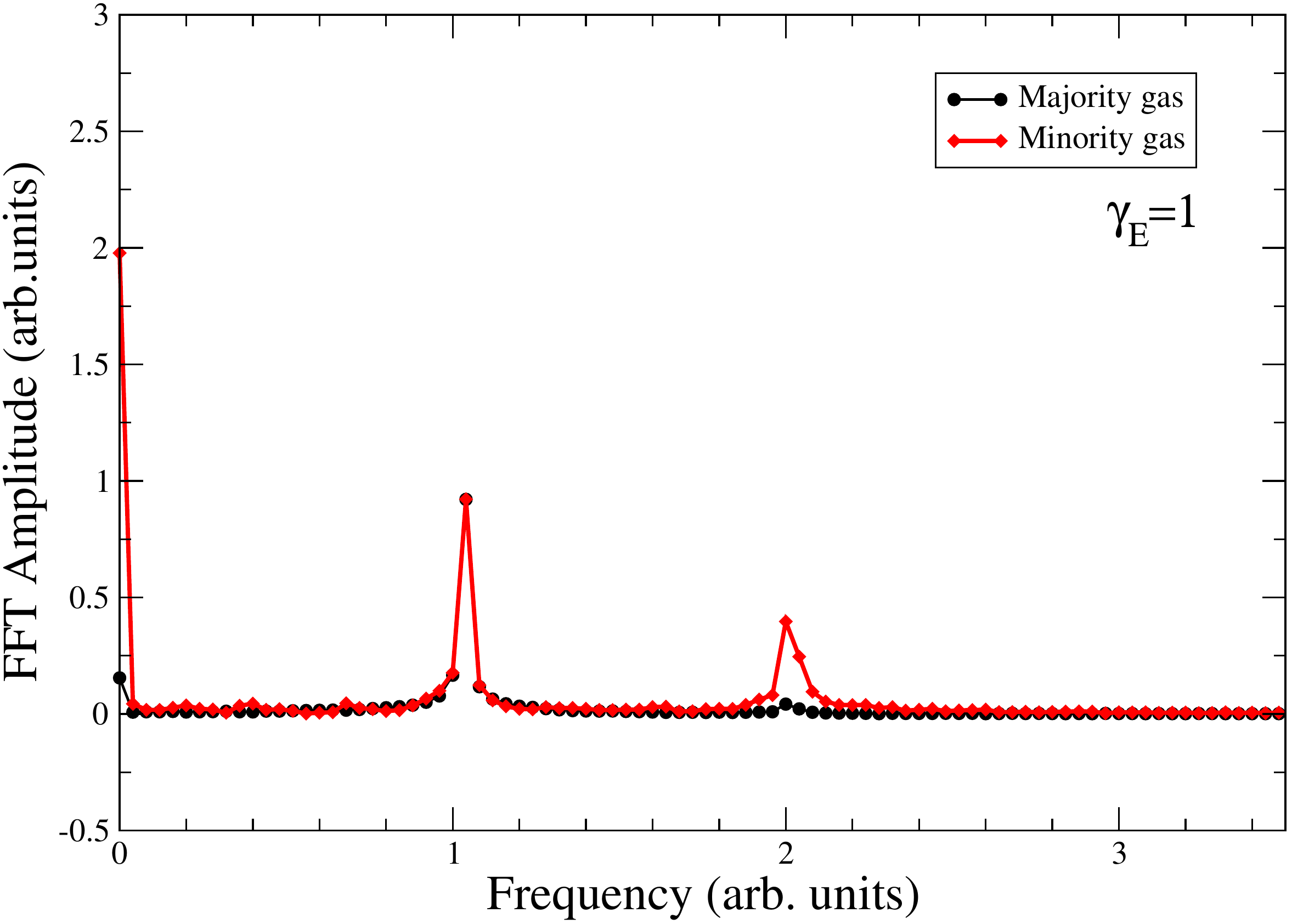} \\
\includegraphics[width=0.42\textwidth,clip=true]{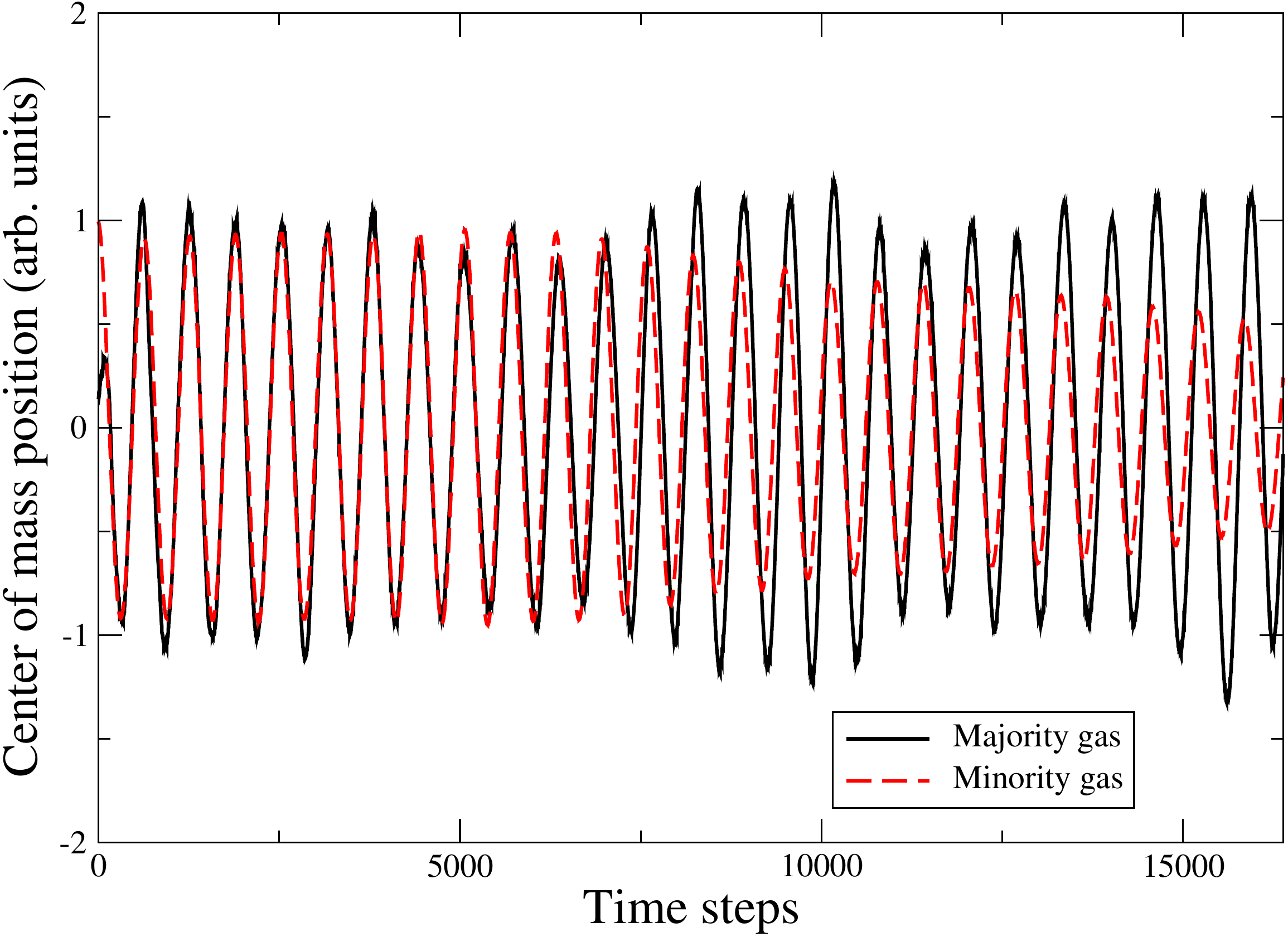} \hspace{0.1in}
\includegraphics[width=0.42\textwidth,clip=true]{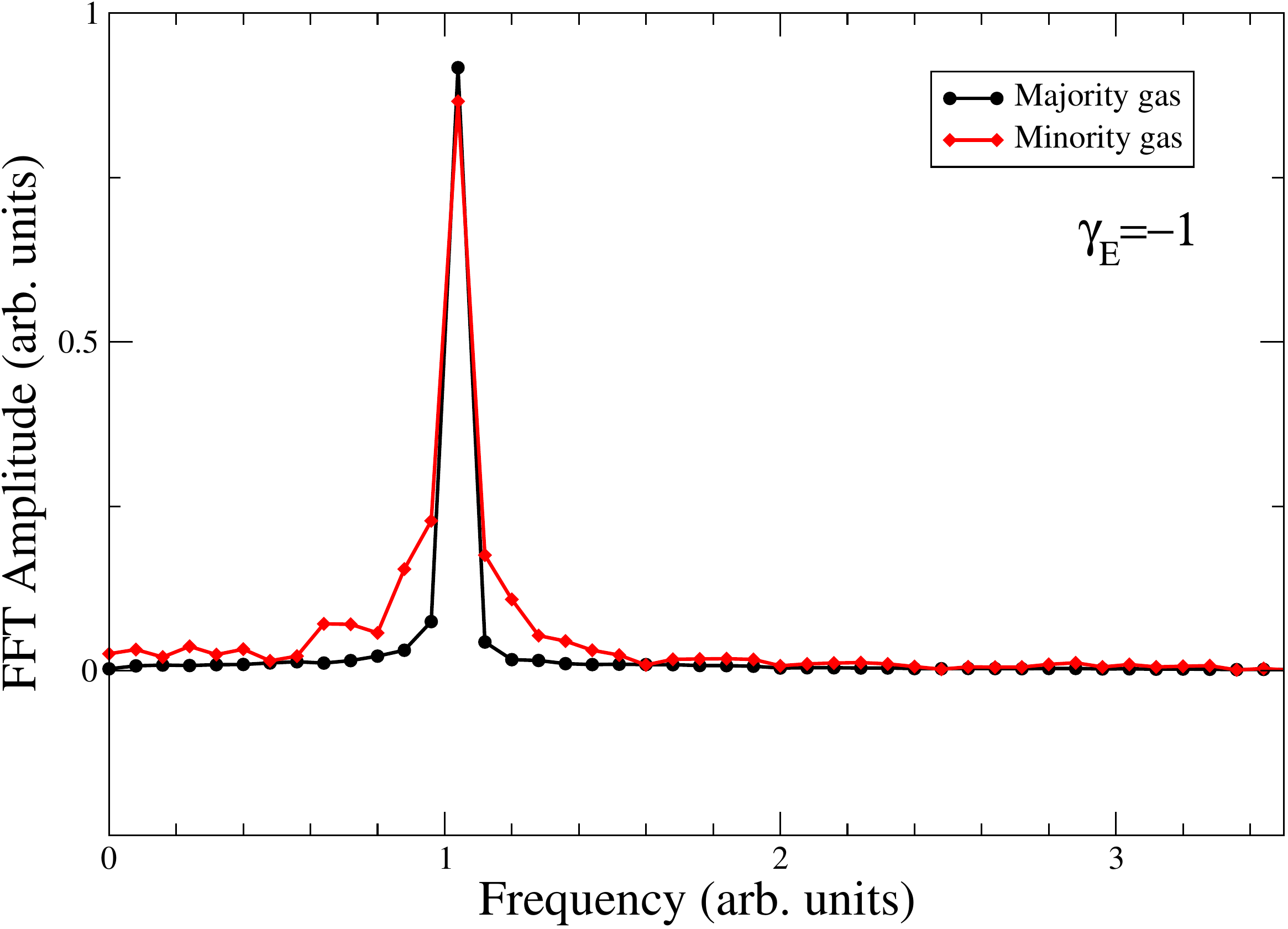}
\caption{(Color online) Case of 1D dynamics with asymmetric particle numbers with 
$N_b=1000, N_p=100$. $\omega=1;\Omega=0.9$, $\beta=1$ and $\lambda=0.1$. The values of 
$\gamma_E$ is specified in the figures. Panels (a), (c) and (e) show the time evolution of 
the centers of mass of the majority cloud (black, continuous lines) and the minority cloud 
(red, dashed lines) while (b), (d) and (f) are the corresponding FFTs (black, dot points, and 
red, diamonds, respectively) over the duration shown in the left panels ($2^{14}$ time steps).}
\label{Fig8}
\end{center}
\end{figure*}

Here we envisage a much simpler strategy in which a single trapping frequency ratio is chosen based on
the initial temperatures. The results are shown in Fig.~\ref{Fig7} where the thermalization curve for 
the optimal frequency ratio is contrasted with those corresponding to smaller and larger ratios. 
The increased efficiency at the optimal ratio is quite apparent. We have not explored this further 
as yet though the usefulness would be most apparent in a well-defined experimental context. 
We do note here that it is crucial to choose this ratio in such a way that the optimal overlap 
occurs at later times, when the initial drop in temperature of the hottest cloud, occurring 
regardless of sophisticated optimizations, has already taken place. 

\section{Number asymmetry and center of mass mode-locking}

Thus far, we have only presented situations where the two species have equal numbers of particles. 
If this condition is upset, interesting phenomena occur as the species with larger number 
of atoms (the 'majority species' thereafter) will lead the thermalization based on the 
dominance of its heat capacity with respect to the species with smaller number of atoms (the 
'minority' species). As long as the nondegenerate limit is considered for small interaction strength, 
this situation is all contained in the Dulong-Petit law with no surprising implications. The case of
strong coupling is not quite so trivial. Here we have to include the ensemble averaged interaction
energy in our estimate of the final thermalized temperature, leading to 
\begin{equation}
T_{eq} = \frac{N_pT_p+N_bT_b}{N_p+N_b} + \frac{\langle E_{\mathrm{int}} \rangle}{(N_p+N_b)k_B} \;.
\end{equation}
In the case of strong coupling and equal numbers, it is clear that the final temperature can exceed
both of the initial values. This was noted both in Ref.~\cite{OnSun} and seen in some of the results here
as well, notably Fig.~\ref{Fig2}(b) and Fig.~\ref{Fig4}(b)/(c). Even if the final temperature lies between 
the initial ones, the interaction energy needs to be included in the calculation of the final temperature, 
which can be quantitatively verified from Fig.~\ref{Fig2}(a). However, where the species numbers are 
strongly asymmetric, $N_p << N_b$, the effect of the interaction term is greatly suppressed, reducing 
it to a weak coupling situation. We note also that minority gases, even for the extreme case of single atoms, 
embedded in majority gases are also crucial for precise and accurate thermometry of 
ultracold gases~\cite{Tuchendler,Sabin,Hohmann}.

A second phenomenon worthy of attention for number-asymmetric clouds is the presence of 'mixing' in 
the frequency spectrum of the oscillations. This may be related to recent measurements on the 
relative motion of two components of an ultracold Fermi-Bose ${}^6$Li-${}^7$Li trapped mixture with 
asymmetric numbers of atoms. Their results indicate a shift in the oscillation frequency of the 
minority species ${}^7$Li while the majority species ${}^6$Li is largely unaffected~\cite{Ferrier,Delehaye}.  
The phenomenon has been quantitatively addressed for the case of a superfluid mixture by considering 
the Bose atoms as impurities immersed in the Fermi superfluid. Adding a mean-field interaction term 
to the external trapping potential, neglecting the back-action of the bosons on the fermions, 
and using a local-density approximation, an effective potential was obtained.  This led
to a trapping frequency of $\tilde{\omega}_b=2 \pi \times 14.97$ Hz, in agreement 
(within two standard deviations) with the measured value of ${\tilde{\omega}_b}^{exp}=
2 \pi \times 15.00(2)$ Hz, and with the experimental value of the trapping frequency 
of ${}^{7}$Li in the absence of ${}^6$Li atoms ${\tilde{\omega}_b}^{exp}=2 \pi \times 15.27(1)$ Hz. 
A second effect was observed above the critical temperatures for superfluidity for both gases, at 
$T \simeq 0.34 T_F, \simeq T_{cB}$, with $T_F$ the Fermi temperature and $T_{cB}$ the critical temperature 
for Bose-Einstein condensation (coinciding with the critical temperature for Bose superfluidity), 
in which ${}^7$Li oscillates at the same frequency of ${}^6$Li (also equaling the damping 
constant)~\cite{Delehaye}. In the same paper, a qualitative description was proposed 
in terms of a ``sort of quantum Zeno effect'' (see details in the supplementary informations of 
Ref~\cite{Delehaye}) related to the violation, due to mean-field effects, of the Kohn 
theorem~\cite{Kohn} for dipole oscillations. 

Our analysis starts by considering two atomic clouds confined around the same location but, in 
general, with different trapping frequencies. The temperatures of the clouds are set to be 
equal, and the center of mass of one of the two clouds is initially offset by a given amount. 
The two clouds have different number of atoms with the mismatch given by $N_2/N_1=10$ such
that we can identify a 'minority' species and a 'majority' species. Their intrinsic trapping 
frequencies are chosen to differ by 10~$\%$. Whenever $\gamma_E \neq 0$ the two clouds will 
interact, modifying the dynamics of their centers of mass. Figure~\ref{Fig8}(a) shows a case 
for small coupling strength ($\gamma_E=10^{-2}$) of repulsive character. A more quantitative 
assessment is achieved by taking the Fast Fourier Transform (FFT) spectrum of the centers of 
mass motion, as shown in Fig.~\ref{Fig8}(b). While the FFT spectrum of the majority cloud 
peaks at a constant value equal to the one in the absence of interactions, the corresponding 
one of the minority cloud has various components, with the highest peak occurring at the 
frequency of the majority cloud. This is effectively a "mode locking" phenomenon. Although 
the analysis is carried out for a 1D system, no further qualitative insights should be gained 
by analyzing a full three-dimensional situation, at least in the nondegenerate regime.

If the interspecies interaction is boosted by two orders of magnitude, the minority cloud center of mass
locks into the frequency of the majority cloud more quickly. Though visible in Fig.~\ref{Fig8}(c), it
is more readily seen in the near coincident peaks in FFT shown in Fig.~\ref{Fig8}(d), even in amplitude, 
with little components at smaller frequencies. Notice that this occurs even if, due to the strong 
repulsive interaction, the centers of mass of the two clouds never intersect each other, as evident 
in Fig.~\ref{Fig8}(c). The FFT analysis also shows an intriguing feature, the presence of a second 
harmonic. This can be explained by the intrinsic nonlinear nature of the interaction term. 
By Taylor-expanding this term in the difference between two coordinates, we notice that next 
to the quadratic term in $\xi_{nm}=(q_n-Q_m)/\lambda$ is a quartic term in the same variable. 
This term is relevant in the strong coupling regime (large $\gamma_E$) and when the expansion 
term $\xi_{nm}$ is large, {\it i.e.} if the average separation between two atoms of the two clouds 
is large and/or the interaction range $\lambda$ is small. Under these circumstances second harmonic 
generation becomes favourable and could result in precision assessment of nonlinearities in the
interaction between species. Notice that strong interactions alone are not a sufficient condition. 
If the interaction is attractive, for the same $\gamma_E$ used in the repulsive case, we see 
from Fig.~\ref{Fig8}(e) and (f) that there is no second harmonic generation in spite of a 
nearly complete overlap between the centers of mass of the two clouds. 

\section{Conclusions}

Motivated by sympathetic cooling of atomic gases, we have extended our exploration of a modified 
one-dimensional Caldeira-Leggett model to higher dimensions in the nondegenerate regime. 
This process isolated some universal characteristics which we argue are largely 
independent of the details of the interspecies interaction. 
The basis for this inference is the transition of the thermalization dynamics from a deterministic regime
to one described by statistical measures. However, the dynamical details are relevant for describing
transients which are important in the design of optimal sympathetic cooling protocols. 
In the process, we also suggest a metric for numerically determining the onset of thermalization which may
be useful for quantifying the efficiency of contrasting strategies. We note that
there are other recent experimental contexts, such as the high-resolution observation of Brownian motion
in optical tweezers~\cite{Raizen},  where the universal scaling aspects of thermalization discussed here may
be directly tested.

We have also explored the dependence of thermalization on the nature of the interaction as well as 
parameters specifically pertinent to higher dimensional dynamics, notably angular momentum. 
It should be noted that we have encountered cases in 3D where, for strong interactions and as a consequence of 
angular momentum mismatch between the two clouds, the onset of thermalization is significantly delayed. 
The principal reason for this behaviour appears to be marginal overlap between the evolved clouds after 
some initial energy exchange. We envision revisiting these situations by using the center of mass dynamics 
of each cloud to verify if this is, in fact, the mechanism. Here, we have illustrated the value of considering 
the center of mass dynamics by identifying a mode-locking mechanism relevant to recently reported experiments 
on Bose-Fermi mixtures in the normal regime. 

In our model, this mode-locking results from nonlinearities in the interspecies interactions. 
In realistic experimental setups, nonlinearities may also arise from inherent aspects of 
the trapping potential which can, in general, enhance the effects due to interactions. 
This is particularly manifest in optical dipole traps \cite{Adams}, which will be the 
focus of future work especially with regard to their thermalization efficiency of cold 
atomic samples in the nondegenerate regime.

\end{document}